\documentclass[aps,prc,twocolumn]{revtex4-2}
\usepackage{graphicx}
\usepackage{amsmath}
\usepackage{color}
\usepackage{array}
\usepackage{tabularx}

\begin{document}
\title{\textit{Ab initio} study of $^7$Li with coupled mass partitions}
\author{Jakub Herko}
\email{jherko@triumf.ca}
\affiliation{TRIUMF, 4004 Wesbrook Mall, Vancouver, British Columbia V6T 2A3, Canada}
\affiliation{Department of Physics and Astronomy, University of Notre Dame, Notre Dame, Indiana 46556-5670, USA}
\affiliation{Lawrence Livermore National Laboratory, 7000 East Ave, Livermore, California 94550, USA}
\author{Konstantinos Kravvaris}
\author{Sofia Quaglioni}
\affiliation{Lawrence Livermore National Laboratory, 7000 East Ave, Livermore, California 94550, USA}
\author{Petr Navr\'atil}
\affiliation{TRIUMF, 4004 Wesbrook Mall, Vancouver, British Columbia V6T 2A3, Canada}
\affiliation{University of Victoria, 3800 Finnerty Road, Victoria, British Columbia V8P 5C2, Canada}
\author{Guillaume Hupin}
\affiliation{Universit\'e Paris-Saclay, CNRS/IN2P3, IJCLab, 91405 Orsay, France}
\author{Mark A. Caprio}
\affiliation{Department of Physics and Astronomy, University of Notre Dame, Notre Dame, Indiana 46556-5670, USA}

\begin{abstract}
\edef\oldrightskip{\the\rightskip}
\begin{description}
\rightskip\oldrightskip\relax
\item[Background]  
Lithium is of broad interest in nuclear astrophysics, fusion energy research, and nuclear technology. From a theoretical perspective, the nucleus $^7$Li presents a remarkable challenge, as its bound and resonant states can exhibit contributions from both the $^4$He + $^3$H cluster configuration and configurations involving a neutron or proton coupled to a $^6$Li or $^6$He core, respectively.
\item[Purpose]   
We aim to achieve a unified \textit{ab initio} description of bound-state and continuum properties of $^7$Li by explicitly including simultaneously the coupled mass/charge partitions $^4$He + $^3$H, $^6$Li + $n$, and $^6$He + $p$. Specifically, we investigate the effect of inter-partition coupling on the spectrum of $^7$Li and calculate cross sections for the $^6$Li($n,p)^6$He, $^6$He($p,n)^6$Li, and $^6$He($p,t)^4$He reactions.
\item[Method]   
We employ the no-core shell model with continuum for the first time in a calculation that couples three mass/charge partitions of the aggregate nucleus $^7$Li, using a chiral nucleon-nucleon interaction as input.
\item[Results]     
The calculated spectrum reproduces all the experimentally observed states of $^7$Li in the correct order and predicts additional resonances. The calculation also reproduces the overall energy dependence of the $^6$Li$(n,p)^6$He cross section. Improved agreement with measured cross sections is obtained after phenomenological adjustment of resonance energies.
\item[Conclusions]
The present results show that coupling the relevant mass/charge partitions is important for a consistent description of the 
$^7$Li spectrum and reaction cross sections, and offers a useful framework for interpreting existing data and guiding future measurements.
\end{description}
\end{abstract}

\maketitle

\section{Introduction}

Lithium is of considerable interest in nuclear astrophysics and nuclear technology. In primordial nucleosynthesis, the predicted abundances of $^{6,7}$Li provide a test of the big-bang model~\cite{Copi1995,Schramm1998,Burles1999,Nollett2000,Nollett2001}. Furthermore, lithium is a core component of tritium-breeding blankets designed for future fusion energy systems such as DEMO~\cite{DEMO}. Natural lithium consists of roughly 95\% $^7$Li, with the rest being $^6$Li. The lighter $^6$Li isotope is particularly important for nuclear technology applications, since the neutron-induced triton production cross section,  $^6\mathrm{Li}(n,^3\mathrm{H})^4\mathrm{He}$, is one of the ten neutron cross section standards  up to 1 MeV of neutron energy~\cite{Carlson2018,CIELO}.

Reactions leading to the compound $^7$Li system have been investigated extensively both experimentally~\cite{Schroder1987,Burzynski1987,Brune1994,Sauvan2001,Park2006,Firestone2016} and theoretically~\cite{Kajino1986,Mertelmeier1986,Csoto2000,Nollett2001,Neff2011,Vasilevsky2012,Solovyev2014,Eraly2016,Solovyev2017,Vorabbi2019,Solovyev2019,Mason2009,Dubovichenko2010,Tursunov2018,Tursunov2021,Mertelmeier1986,Csoto2000,Vasilevsky2012}. 
In particular, the microscopic studies of Refs.~\cite{Kajino1986,Mertelmeier1986,Csoto2000,Nollett2001,Neff2011,Vasilevsky2012,Solovyev2014,Eraly2016,Solovyev2017,Vorabbi2019,Solovyev2019} address the $^4$He + $^3$H reaction, while Refs.~\cite{Mason2009,Dubovichenko2010,Tursunov2018,Tursunov2021} treat the same process within potential models. The studies of Refs.~\cite{Mertelmeier1986,Csoto2000,Vasilevsky2012} also consider the $^6$Li + $n$ channel. An \textit{ab initio} description based on realistic inter-nucleon interactions was achieved within the no-core shell model with continuum (NCSMC)~\cite{Eraly2016,Vorabbi2019}. In Ref.~\cite{Eraly2016}, the focus was on the radiative capture of triton by an $\alpha$-particle producing $^7$Li or $^4$He($^3$H,$\gamma$)$^7$Li, while Ref.~\cite{Vorabbi2019} extended the investigation by also considering the remaining relevant binary mass/charge partitions (or, simply, partitions), namely $^6$Li + $n$ and $^6$He + $p$, albeit not simultaneously, but as separate calculations. That study provided a good description of the spectrum of $^7$Li, reproducing all the experimentally established states in the correct order and predicting additional resonances not yet observed experimentally. 

One of the resonances predicted in Ref.~\cite{Vorabbi2019} is an $S$-wave $1/2^+$ resonance located just above the $^6$He + $p$ threshold and identified in the $^6$He + $p$ partition. Motivated by this prediction, the $^7$Li($d,^3$He)$^6$He measurement reported in  Ref.~\cite{Dronchi2023} searched for evidence of this  in the $^6$He + $p$ channel, but no such signature was observed. As noted above, however, the calculations of Ref.~\cite{Vorabbi2019} treated the relevant partitions separately and neglected the coupling between them, which was identified as a possible origin of the discrepancy between the prediction and the experimental findings~\cite{Dronchi2023}.

The need for a coupled treatment of $^4$He+$^3$H, $^6$Li+$n$, and $^6$He+$p$ partitions can be understood from the threshold structure of the relevant $^7$Li binary decay channels. The partition $^4$He + $^3$H is the most important for description of $^7$Li, because the $^4$He + $^3$H breakup has the lowest threshold and therefore all resonances are affected by it. The $^6$Li + $n$ threshold lies at higher energy, followed by the $^6$He + $p$ threshold.  Resonances lying above the $^6$Li + $n$ ($^6$He + $p$) threshold can decay by emission of a neutron (proton). A proper description of all the full resonance spectrum therefore requires all three partitions to be coupled within a single calculation. Such coupling is also essential for the description of charge-exchange and nucleon-transfer reactions.

In the present work, we investigate the properties of the $^7$Li system within the NCSMC, taking into account the partitions $^4$He + $^3$H, $^6$Li + $n$, and $^6$He + $p$ in a single coupled-channels calculation. Enabled by recent computational developments~\cite{Kravvaris2024}, this is the first NCSMC study to couple partitions simultaneously. This framework allows us to assess the impact of inter-partition coupling and to calculate cross sections for the  $^6$Li($n,p$)$^6$He, $^6$He($p,n$)$^6$Li, and $^6$He($p,t)^4$He reactions. The $^6$Li($n,p$)$^6$He cross section is compared with available experimental data, while the other results provide predictions for a planned experiment at the Canadian particle accelerator center TRIUMF. That measurement will probe resonances in $^7$Li above the $^6$He + $p$ threshold by directing a $^6$He beam onto a target with hydrogen, thereby studying $^6$He + $p$ reactions in inverse kinematics. To provide additional benchmarks for that experiment, we also calculate differential cross sections for elastic proton scattering on $^6$He.

Section~\ref{sec2} provides a brief overview of the NCSMC formalism, and the results of our calculations are presented in Section~\ref{sec3}. In section~\ref{sec:bound}, we discuss the bound-state energies. Section~\ref{sec:continuum} presents  phase shifts, resonance energies and widths, and cross sections for neutron  elastic scattering on $^6$Li, the charge-exchange reactions $^6$Li($n,p)^6$He and $^6$He($p,n)^6$Li, proton elastic scattering on $^6$He, and the nucleon-transfer reaction $^6$He($p,t)^4$He. In Section~\ref{sec:pheno}, we introduce phenomenological adjustments to resonance energies to improve agreement between the calculated and experimental  cross sections for the $^6$Li($n,p)^6$He reaction and neutron elastic scattering on $^6$Li; the impact of these adjustments on other cross sections is also examined. Finally, section~\ref{sec4} summarizes our conclusions.

\section{Methods}
\label{sec2}

In this work the many-body system is considered to consist of point-like, non-relativistic, mutually interacting nucleons. 
It is described by an intrinsic Hamiltonian $H$ consisting of the intrinsic kinetic energy operator and a nuclear interaction that acts between pairs of nucleons (nucleon-nucleon or $NN$ interaction). 

Here we employ the chiral $NN$ interaction at next-to-next-to-next-to-leading order (or, N$^3$LO) of Refs.~\cite{Entem2003,Machleidt2011}. To accelerate convergence with increasing model-space size, we soften the $NN$ interaction using a similarity-renormalization-group (SRG) transformation~\cite{Bogner2007,JurgensonA,JurgensonB} with flow parameter $\lambda_{\rm SRG}=2.15$ fm$^{-1}$. This choice is consistent with Refs.~\cite{Eraly2016,Vorabbi2019}, where it was shown to reproduce the $^4$He+$^3$H threshold without inclusion of three-body forces. In this work, we neglect SRG-induced three-body and higher-body forces; their effect  will be investigated in future work.

To obtain many-body static solutions for this Hamiltonian, we employ the no-core shell model (NCSM) framework ~\cite{Navratil2000,Barrett2013}. In this approach, the Schr\"odinger equation $H|\Psi\rangle=E|\Psi\rangle$ for a system of interacting nucleons is solved by representing the Hamiltonian in a basis of Slater determinants constructed from spherical harmonic oscillator (HO) single-particle wave functions. The resulting matrix representation is sparse, with the eigenpairs $(E_n,|\Psi_n\rangle)$ representing the solutions of the Schr\"odinger equation. These eigenpairs can be computed, for example, using the Lanczos method~\cite{Lanczos1950}, even for matrices with dimension upwards of $10^{10}$~\cite{McCoy2024}. The HO single-particle basis states are characterized by oscillator frequency $\Omega$, which corresponds to an oscillator energy quantum of $\hbar\Omega$. The size of the many-body Slater determinant basis is determined by the maximum number of excitation quanta, $N_{\rm max}$.

Although the spectrum of $^7$Li can be described reasonably well within the NCSM, the calculation of reaction observables requires a framework that also accounts for continuum degrees of freedom. For this purpose,  we employ the NCSMC~\cite{Baroni2013L,Baroni2013C,Navratil2016,Quaglioni2018}, which provides a unified description of bound and scattering states. In the NCSMC, the wave function of the $A$-nucleon system is expanded on a basis consisting of NCSM eigenstates of the intrinsic $A$-nucleon Hamiltonian and continuous microscopic binary-cluster states within the resonating-group method (RGM)~\cite{Tang1978}. In these RGM-type states, the intrinsic wave functions of the clusters are eigenstates of the same intrinsic Hamiltonian $H$ as the aggregate system, also calculated in the NCSM framework~\cite{Quaglioni2008,Quaglioni2009,Navratil2016}. 

In a compact form, the wave function of the reacting $A$-body system for specific total angular momentum and parity $J^\pi$ is written as
\begin{equation}\label{eq1}
|\Psi^{J^{\pi}}\rangle=\sum_{\lambda}c_{\lambda}^{J^{\pi}}|A\lambda J^{\pi}\rangle+\sum_{\nu}\int_{0}^{+\infty}\textrm{d}r\,r^{2}\frac{\gamma_{\nu}^{J^{\pi}}(r)}{r}\mathcal{A}_{\nu}|\Phi_{\nu r}^{J^{\pi}}\rangle,
\end{equation}
where $|A\lambda J^{\pi}\rangle$ denotes the NCSM eigenstates of the $A$-body system, labeled by the index $\lambda$, with expansion coefficients $c_{\lambda}^{J^{\pi}}$; $|\Phi_{\nu r}^{J^{\pi}}\rangle$ denotes the continuous binary-cluster RGM basis states, labeled by the channel index $\nu$, with continuous expansion amplitudes $\gamma_{\nu}^{J^{\pi}}(r)$; $r$ is a parameter coordinate, which plays the role of the relative distance between the centers of mass of the clusters; and $\mathcal{A}_{\nu}$ is the inter-cluster  antisymmetrization operator, which accounts for exchanges of identical nucleons between the two clusters. The isospin $T$ of the $A$-body system does not appear in Eq.~(\ref{eq1}), because--in the present calculations involving charge exchange between partitions--the isospins of the clusters are not coupled in the RGM basis states. Nevertheless, $T$ remains an approximate quantum number and is used later in the paper to label the NCSMC eigenstates of $^7$Li.

We note that the channel index $\nu=\{\ell\, s\, a\, z\, \lambda_1\, J_1^{\pi_1}\, \lambda_2\, J_2^{\pi_2}\}$ contains not only the asymptotic quantum numbers, namely the relative angular momentum $\ell$ and channel spin $s$, but also the mass and charge partition of the system into binary clusters. Here, $a$ and $z$ denote the mass and charge of the lighter cluster, respectively, while $\lambda_1, \lambda_2$ label the NCSM states of the clusters with angular momenta and parities $J_1^{\pi_1}$ and $ J_2^{\pi_2}$. This structure allows different mass/charge reaction channels to be incorporated within the same formalism on an equal footing.

In the present application, the aggregate system is $^{7}$Li and the partition entering the channel index $\nu$ corresponds to one of three binary fragmentations: $^4$He + $^3$H, $^6$Li + $n$, and $^6$He + $p$. These three partitions are included simultaneously in a single coupled-channels calculation, thus enabling the extraction of both charge-exchange and nucleon-transfer reaction observables. The corresponding binary-cluster RGM basis states take the form
\begin{multline}\label{bin}
|\Phi_{\nu r}^{J^{\pi}}\rangle=\left[\left(\left|X\lambda_1J_1^{\pi_1}
\right\rangle\left|Y\frac{1}{2}^{+}\right\rangle\right)^{(s)}Y_{\ell}(\hat{r}_{12})\right]^{(J^{\pi})}\\
\times\frac{\delta(r-r_{12})}{rr_{12}},
\end{multline}
where $\left|X\lambda_1J_1^{\pi_1}\right\rangle$ is the $\lambda_1$-th NCSM eigenstate of $X={^4\textnormal{He}}$, ${^6\textnormal{Li}}$, or ${^6\textnormal{He}}$ with angular momentum $J_1$ and parity $\pi_1$, and $\big|Y\frac{1}{2}^{+}\big\rangle$ is the intrinsic wave function of the ground state of triton, neutron, or proton ($Y={^3}\textnormal{H}$, $n$, or $p$), respectively. Furthermore, $\vec{r}_{12}=r_{12}\hat{r}_{12}$ is the relative displacement vector between the centers of mass of the two clusters, and $Y_{\ell}(\hat{r}_{12})$ denote spherical harmonics. The total angular momenta $J_1$ and   $\frac{1}{2}$ of the clusters $X$ and $Y$ are coupled  to the channel spin $s$, which in turn is coupled with the relative orbital angular momentum $\ell$ to the total angular momentum $J$. Therefore, for the channels considered here, the partition, along with the quantum numbers $\lambda_1, J_1^{\pi_1}, s$, and $\ell$, fully specifies the channel index $\nu$. The NCSM eigenstates of $X$ also carry an approximate isospin $T_1$ quantum number, which we use later in the paper to label these states. As already mentioned, this isospin is not coupled with the isospin of $Y$ to the total isospin $T$.

The discrete expansion coefficients $c_{\lambda}^{J^{\pi}}$ and the continuous amplitudes $\gamma_{\nu}^{J^{\pi}}(r)$, which represent the relative motion of the clusters, are obtained by solving the NCSMC equations [see Eq.~(3) in Ref.~\cite{Baroni2013L}] using an extention of the $R$-matrix method on a Lagrange mesh~\cite{Baroni2013C}. The NCSMC equations can be solved for either bound or scattering states by adopting the appropriate asymptotic ansatz for the relative-motion amplitude $\gamma_{\nu}^{J^{\pi}}(r)$. For scattering states, the asymptotic behavior of $\gamma_{\nu}^{J^{\pi}}(r)$ determines the scattering matrix, from which phase shifts and cross sections can be obtained.

To evaluate the localized contributions to the integral kernels entering the NCSMC equations, the Dirac $\delta$-function in the binary-cluster states~(\ref{bin}) is expanded in terms of radial HO wave functions~\cite{Quaglioni2009,Navratil2016}. These functions are defined with the same frequency $\Omega$ used in the NCSM calculations for $^4$He, $^3$H, $^6$Li, $^6$He, and $^7$Li, and the expansion is truncated at a maximum number of HO quanta consistent with the corresponding NCSM model spaces.

\section{Results}
\label{sec3}

In this section we present the results of our NCSMC calculations for the $^{7}$Li system, including bound state energies, phase shifts, resonance energies and widths, and cross sections for the reactions $^6$Li($n,n)^6$Li, $^6$Li($n,p)^6$He, $^6$He($p,p)^6$He, $^6$He($p,n)^6$Li, and $^6$He($p,t)^4$He. We compare our results with available experimental data and provide predictions for resonances that have not yet been observed experimentally.

\subsection{NCSM eigenstates and model space choices}
\label{sec:model_space}
In the NCSM calculations of the eigenstates entering the NCSMC basis,  we used HO model spaces up to $N_{\rm max} =12$ for $^4$He and $^3$H, up to $N_{\rm max} =10$ for $^6$Li, $^6$He, and the negative-parity states of $^7$Li, and up to $N_{\rm max} =11$ for the positive-parity states of $^7$Li. These choices reflect current computational limitations of NCSMC calculations. Although the NCSM allows calculations in larger model spaces, the computation of reaction observables remains considerably more demanding from a technical and computational standpoint.

\begin{figure}
\includegraphics[width=\linewidth]{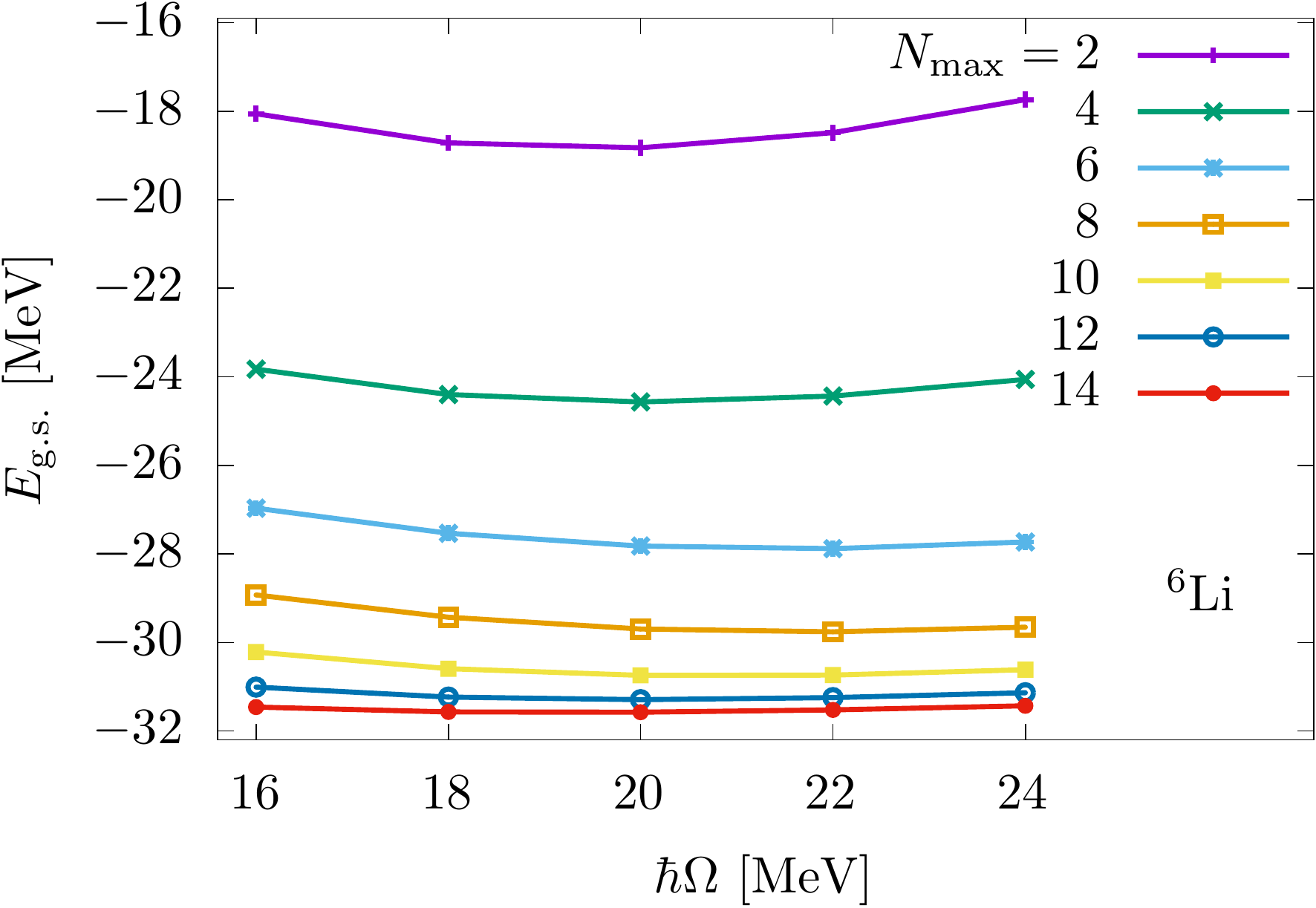}
\caption{Ground state ($J_1^{\pi_1} T_1=1^+0$) energy of $^6$Li calculated within the NCSM as a function of $\hbar\Omega$  for different values of $N_{\rm max}$.}
\label{6Lihw}
\end{figure}

For this work, we adopt $\hbar\Omega=20$ MeV, which is close to the value for which the energies of the NCSM eigenstates of $^6$Li, $^6$He, and $^3$H reach their minimum. This behavior is illustrated in Fig.~\ref{6Lihw}: as the model space increases from $N_{\rm max}=2$ to 14, the $\hbar\Omega$ dependence of the ground-state energy of $^6$Li flattens, showing a clear approach to convergence by $N_{\rm max}=12$. A similar behavior is found for the other $^6$Li states and for the $^6$He and $^3$H states included in the present NCSMC calculations. For $^4$He, the minimum occurs near $\hbar\Omega=24$ MeV, but the $\hbar\Omega$ dependence is very weak. 

In the present work, the $^4$He+$^3$H, $^6$Li+$n$, and $^6$He+$p$ partitions  are included in a single calculation that explicitly accounts for their coupling. However, to gauge the effect of such coupling, we also performed calculations taking into account the three partitions separately, as in the previous work of Ref.~\cite{Vorabbi2019} (which included fewer NCSM eigenstates of the compound nucleus $^7$Li). The same values of $\hbar\Omega$ and $N_{\rm max}$ were used in Ref.~\cite{Vorabbi2019}, allowing for a direct comparison with the present calculation.

We include the lowest twelve negative-parity and the lowest six positive-parity NCSM eigenstates of the compound nucleus $^{7}$Li, which encompass all experimentally observed states of $^7$Li. For $^4$He and $^3$H, we include only their $0^+0$ and $1/2^+1/2$ ground states, respectively. For $^6$Li, we include four NCSM eigenstates with $J_1^{\pi_1}T_1=1^+0,\,3^+0,\,0^+1,\,2^+1$, while for $^6$He we include two states with $J_1^{\pi_1}T_1=0^+1,\,2^+1$. These correspond to the experimentally observed bound states and narrow resonances, and include isobaric analog states.

\begin{figure}
\includegraphics[width=\linewidth]{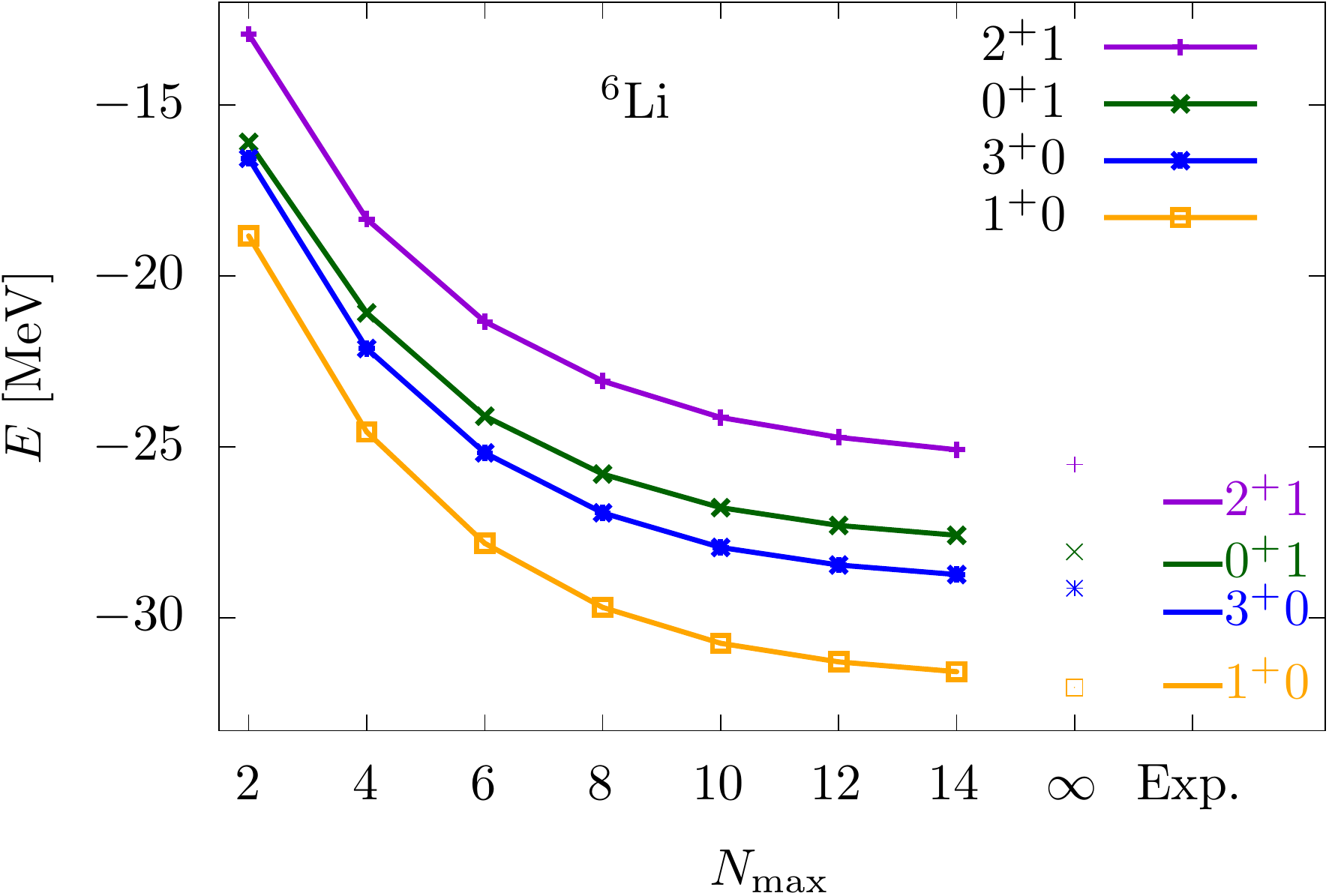}
\caption{The total energies of the NCSM eigenstates of $^6$Li taken into account in the NCSMC calculations as functions of $N_{\rm max}$ for $\hbar\Omega=20$ MeV. Angular momenta, parities, and isospins $J_1^{\pi_1}T_1$ of the states are indicated. Experimental energies are also shown.}
\label{6Liconv}
\end{figure}

The absolute energies of the four NCSM eigenstates of $^6$Li exhibit a clear trend toward convergence with increasing $N_{\rm max}$, as shown in Fig~\ref{6Liconv}. At $N_{\rm max}=10$, which is the value used in the NCSMC calculations, the energies are not yet fully converged. To estimate the converged values, we extrapolate the calculated energies to $N_{\rm max}\to\infty$ using the fit function $E=a\,e^{-bN_{\rm max}}+E_{\infty}$, where $a, b, E_{\infty}$ are free parameters~\cite{Forssen2008,Bogner2008}. Here, $E_{\infty}$ denotes the extrapolated energy in the infinite-space limit and is indicated in the
figures by the horizontal axis label $\infty$. The largest difference between the energy calculated for $N_{\rm max}=14$ and the extrapolated energy is 0.48 MeV or 1.7\% for the 0$^+$1 state. The largest difference between the extrapolated and experimental energies is 1.11 MeV or 4.2\% for the 2$^+$1 state. The convergence behavior of the two NCSM eigenstates of $^6$He is similar, while for the lighter clusters $^4$He and $^3$H it is more favorable.

The ground-state energies and matter root-mean-square radii of $^4$He, $^3$H, $^6$Li, $^6$He, and $^7$Li calculated for the greatest $N_{\rm max}$ used in the NCSMC calculations ($N_{\rm max}=12$ for $^4$He and $^3$H and $N_{\rm max}=10$ for the other nuclei) are shown in Tables~\ref{tab-1} and~\ref{tabr}, respectively. Values calculated using $N_{\rm max}=16$ for $^4$He and $^3$H, $N_{\rm max}=14$ for $^6$Li and $^6$He, and $N_{\rm max}=12$ for $^7$Li are shown as well along with energies extrapolated to $N_{\rm max}\to\infty$ and experimental values. The uncertainty of each extrapolated energy was estimated by performing the fit for $\hbar\Omega=16,18,20,22,24$ MeV and including or excluding the $N_{\rm max}=2,4$ values. The extrapolated energy for $^3$H is the energy calculated for $N_{\rm max}=36$, which is fully converged and thus has no uncertainty. Except this nucleus, the extrapolated energies agree with experiment within the quoted uncertainties.

\begin{table}
\caption{Total NCSM energies $E$ (in MeV) of the ground states of $^4$He, $^3$H, $^6$Li, $^6$He, and $^7$Li calculated for the greatest $N_{\rm max}$ used in the NCSMC calculations ($N_{\rm max}=12$ for $^4$He and $^3$H and $N_{\rm max}=10$ for the other nuclei). Values $\tilde{E}$ calculated using $N_{\rm max}=16$ for $^4$He and $^3$H, $N_{\rm max}=14$ for $^6$Li and $^6$He, and $N_{\rm max}=12$ for $^7$Li are shown as well along with values $E_{\infty}$ extrapolated to $N_{\rm max}\to\infty$ and experimental values.}
\label{tab-1}
\begin{tabularx}{\columnwidth}{*6{>{\centering\arraybackslash}X}}
\hline
\hline
nucleus & $E$ & $\tilde{E}$ & \multicolumn{2}{c}{$E_{\infty}$} & Exp. \\ \hline
$^4$He & $-27.97$ & $-28.02$ & \multicolumn{2}{c}{$-28.12\pm0.42$} & $-28.30$ \\
$^3$H & $-8.24$ & $-8.29$ & \multicolumn{2}{c}{$-8.30$} & $-8.48$ \\
$^6$Li & $-30.74$ & $-31.57$ & \multicolumn{2}{c}{$-32.04\pm0.36$} & $-31.99$ \\
$^6$He & $-27.66$ & $-28.44$ & \multicolumn{2}{c}{$-28.91\pm0.68$} & $-29.27$ \\
$^7$Li & $-38.01$ & $-38.69$ & \multicolumn{2}{c}{$-39.56\pm0.79$} & $-39.25$ \\
\hline
 \hline
\end{tabularx}
\end{table}

Table~\ref{tabr} shows that the radius calculated within the NCSM with the HO basis significantly increases from $^4$He to $^6$He, indicating the onset of halo structure, and, as shown in Ref.~\cite{Caprio2014}, the radius then remains essentially unchanged for $^8$He. This indicates that the description of the halo nucleus $^6$He within the NCSM is reasonable. A better description would be achieved by taking into account the $^4$He + $n$ + $n$ cluster structure, as done within the NCSMC in Refs.~\cite{Redondo2016,Quaglioni2018}, where $^6$He is the aggregate nucleus. However, such description of a cluster is out of the scope of the current capabilities of the NCSMC framework.

\begin{table}
\caption{Matter root-mean-square radii $R$ (in fm) for the ground states of $^4$He, $^3$H, $^6$Li, $^6$He, and $^7$Li calculated within the NCSM for the greatest $N_{\rm max}$ used in the NCSMC calculations ($N_{\rm max}=12$ for $^4$He and $^3$H and $N_{\rm max}=10$ for the other nuclei). Values $\tilde{R}$ calculated using $N_{\rm max}=16$ for $^4$He and $^3$H, $N_{\rm max}=14$ for $^6$Li and $^6$He, and $N_{\rm max}=12$ for $^7$Li are shown as well along with experimental values.}
\label{tabr}
\begin{tabularx}{\columnwidth}{*4{>{\centering\arraybackslash}X}}
\hline
\hline
nucleus & $R$ & $\tilde{R}$ & Exp. \\ \hline
$^4$He & $1.46$ & $1.46$ & $1.49\pm0.03$~\cite{Egelhof2002} \\
$^3$H & $1.68$ & $1.70$ &  \\
$^6$Li & $2.08$ & $2.15$ & $2.45\pm0.07$~\cite{Egelhof2002} \\
$^6$He & $2.10$ & $2.18$ & $2.30\pm0.07$~\cite{Egelhof2002} \\
$^7$Li & $2.14$ & $2.17$ & $2.34\pm0.04$~\cite{Tanihata1988} \\
\hline
 \hline
\end{tabularx}
\end{table}

\subsection{Bound state spectra}
\label{sec:bound}

\begin{figure}
\includegraphics[width=\linewidth]{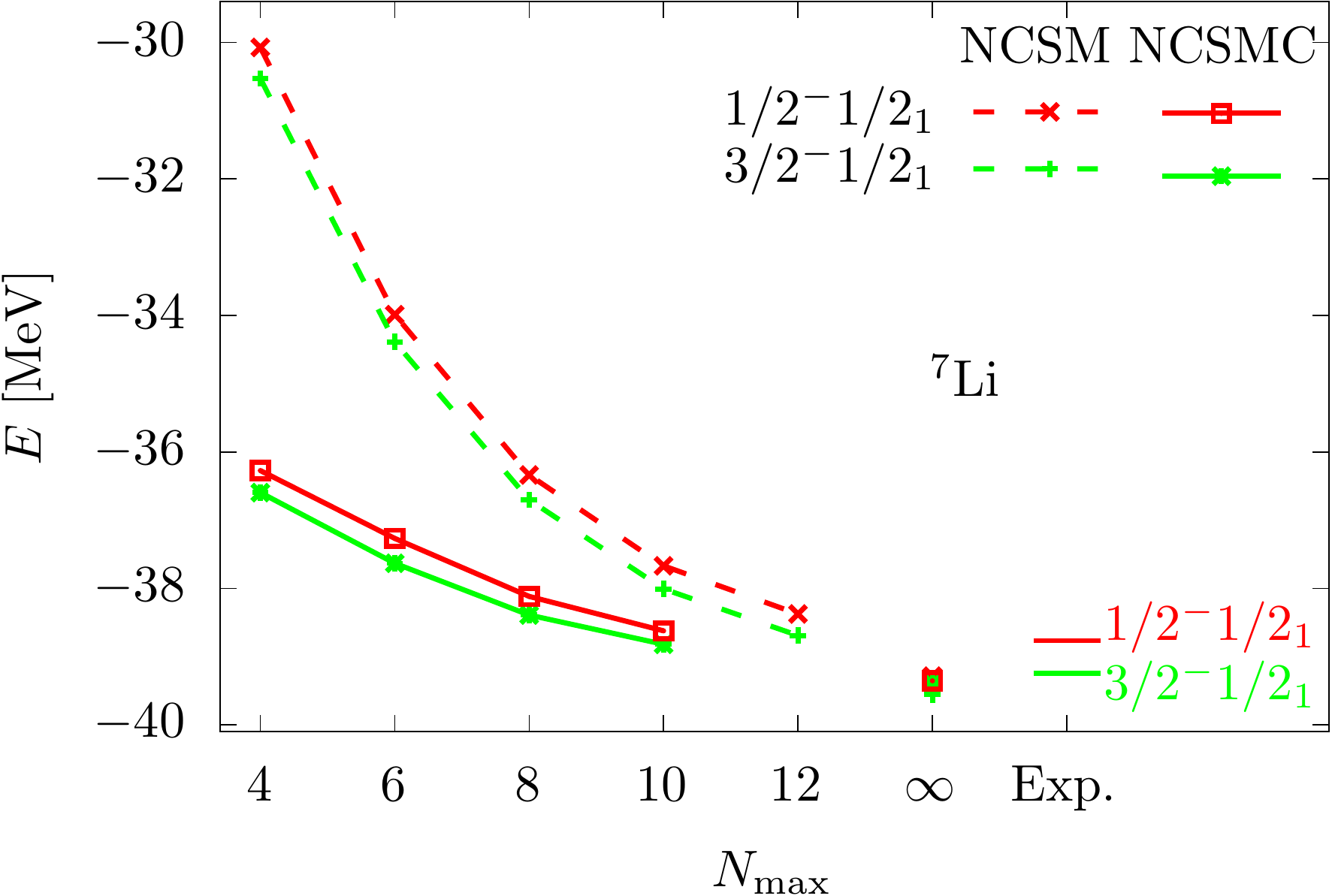}
\caption{Observed exponential convergence for the low-lying (bound) spectrum of $^7$Li with the NCSM and NCSMC compared to experiment. The NCSMC calculations produce states that are more bound, while following an overall convergence pattern similar to that of the NCSM. Note that the extrapolated NCSM and NCSMC values are almost the same.}
\label{7Liconv}
\end{figure}

We begin with the NCSMC results for the bound states of $^7$Li, using the NCSM results as a reference. The convergence of the total energy of the bound states of $^7$Li calculated within the NCSM and NCSMC with respect to $N_{\rm max}$ is shown in Fig.~\ref{7Liconv}. At a given value of $N_{\rm max}$, the NCSMC calculations provide additional binding relative to the NCSM, bringing the energies closer to the infinite-model-space extrapolation, although both approaches converge to the same result within their uncertainties. For the NCSM results, the largest difference between the energy calculated for $N_{\rm max}=12$ and the extrapolated energy is 0.91 MeV or 2.3\% for the $J=1/2$ state, and  between the extrapolated and experimental energies is 0.52 MeV or 1.3\% for the same state. Conversely, for the NCSMC results, the largest difference between the extrapolated and experimental energies is 0.59 MeV (1.5\%) for the $J=1/2$ state.

The total NCSM energy of the ground state of $^7$Li calculated for $N_{\rm max}=12$ and extrapolated to $N_{\rm max}\to\infty$ is shown in Table~\ref{tab-1} along with the corresponding experimental value. The extrapolated value agrees with the experimental value within the uncertainty.

\begin{figure}
\includegraphics[width=\linewidth]{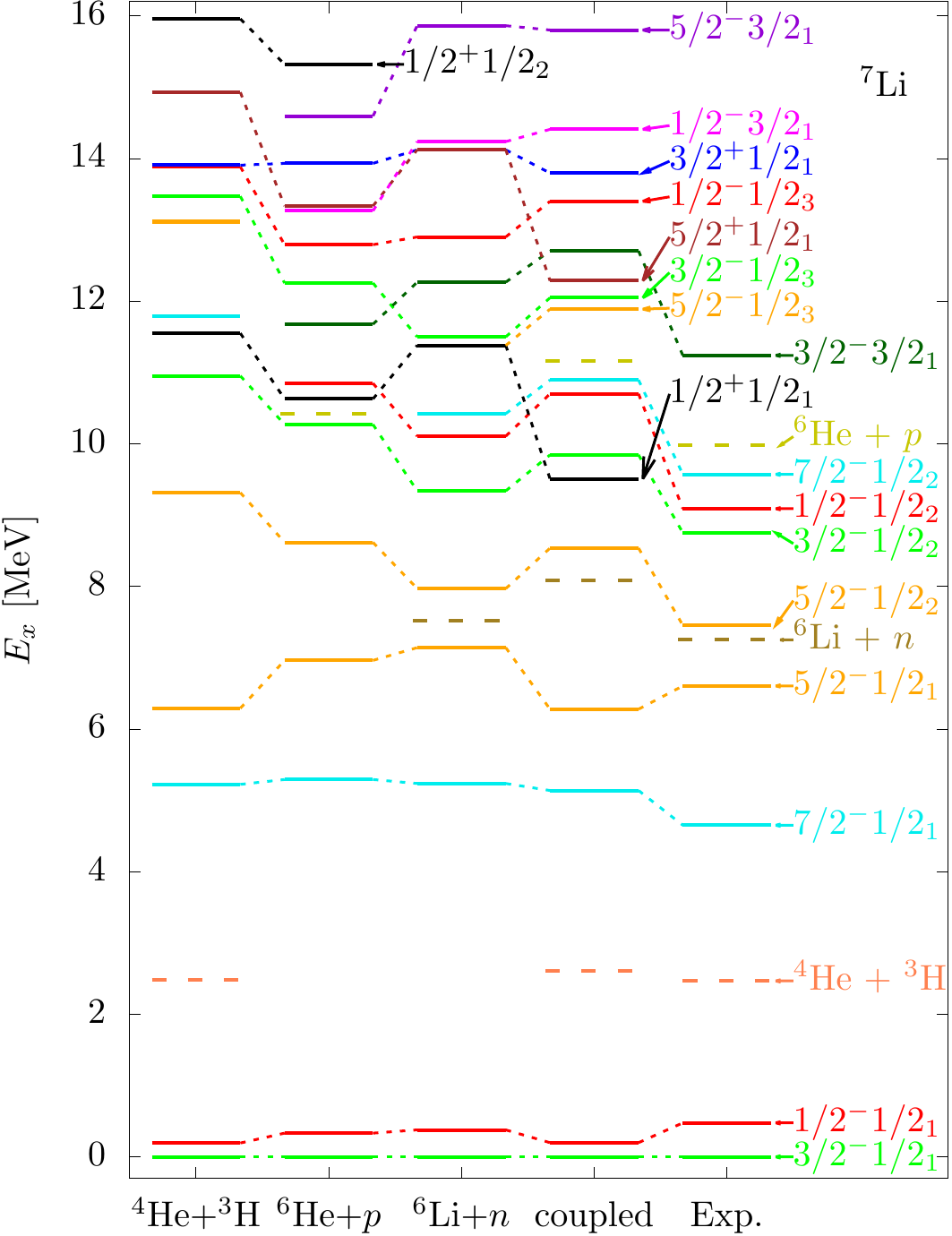}
\caption{The spectra of $^7$Li calculated within the NCSMC along with experimental data. The spectrum obtained in the coupled-partition calculation is shown along with the spectra obtained in the calculations taking into account the three partitions separately. The dashed lines denote the $^4$He + $^3$H, $^6$Li + $n$, and $^6$He + $p$ thresholds.}
\label{fig4}
\end{figure}

The energies of the bound states of $^7$Li relative to the $^4$He + $^3$H threshold calculated within the NCSMC for the largest value of $N_{\rm max}$ are shown in Table~\ref{tab0} along with the corresponding experimental values. The excitation energy of the $1/2^-1/2_1$ state is also shown in Fig.~\ref{fig4}, where we present the spectra of $^7$Li calculated within the NCSMC for the largest value of $N_{\rm max}$  and compare them with the experimental spectrum~\cite{Tilley2002}. The results of the NCSMC calculations including the three partitions separately are also shown in Fig.~\ref{fig4}.

\begin{table}
\caption{NCSMC energies (in MeV) of the bound states of $^7$Li relative to the $^4$He + $^3$H threshold calculated for the greatest value of $N_{\rm max}$. For this value of $N_{\rm max}$ the threshold energy is $-36.21$ MeV. Experimental values are shown as well.}
\label{tab0}
\begin{tabularx}{\columnwidth}{*3{>{\centering\arraybackslash}X}}
\hline
\hline
$J^{\pi}T_i$ & NCSMC & Exp. \\ \hline
$3/2^-1/2_1$ & $-2.61$ & $-2.47$ \\
$1/2^-1/2_1$ & $-2.41$ & $-1.99$ \\
\hline
 \hline
\end{tabularx}
\end{table}

\subsection{Continuum}
\label{sec:continuum}

\subsubsection{Resonance analysis}
\label{sec:res}

\begin{figure}
\includegraphics[width=\linewidth]{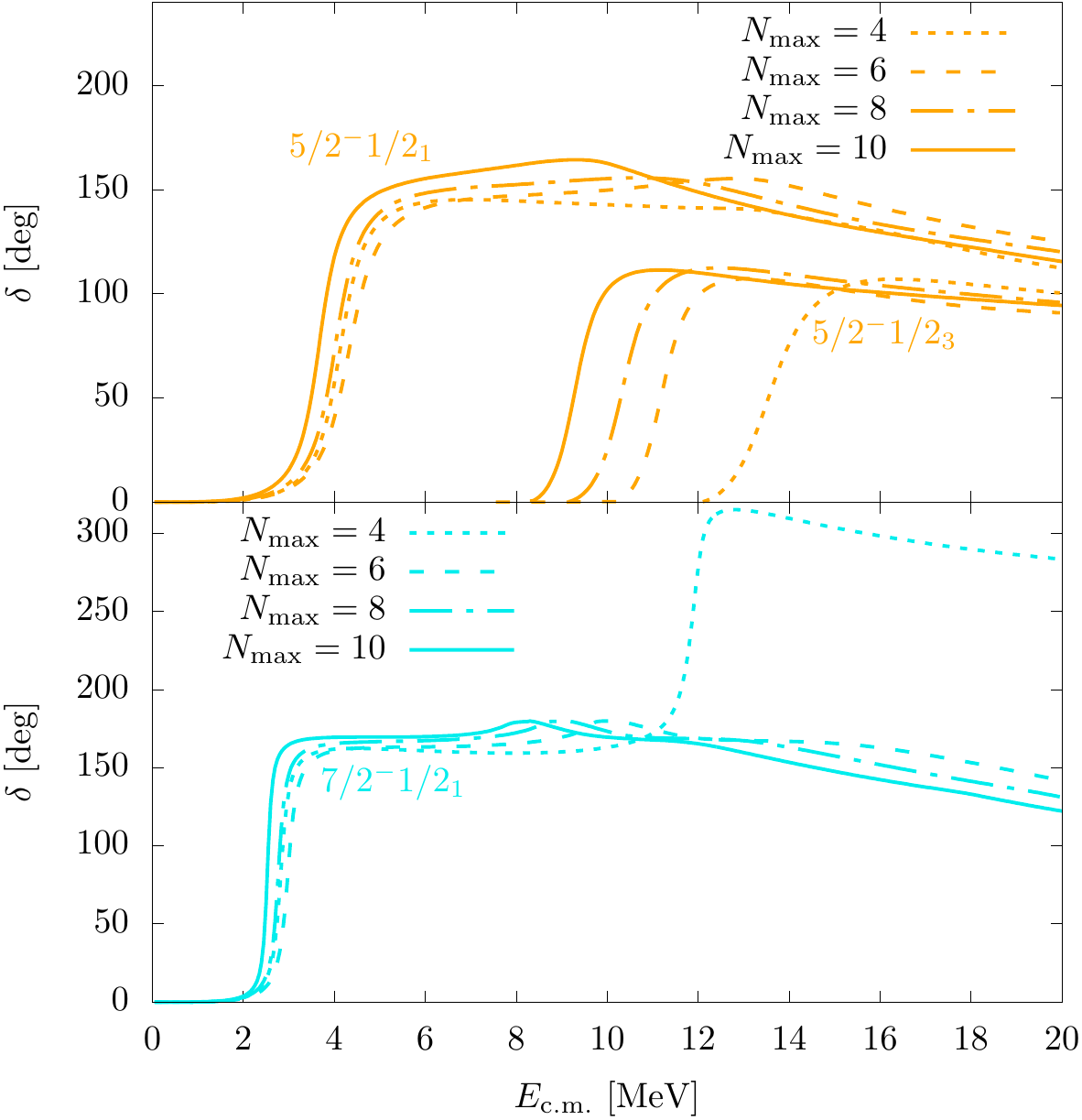}
\caption{The eigenphase shifts of the $5/2^-1/2_1$, $5/2^-1/2_3$ (top), and $7/2^-1/2_1$ (bottom) resonances in $^7$Li as functions of the energy in the center-of-mass frame relative to the $^4$He + $^3$H threshold calculated for different values of $N_{\rm max}$. Convergence of other eigenphase shifts in the negative-parity $J^{\pi}T$ channels is similar to that of the eigenphase shift of the $5/2^-1/2_3$ resonance. The $7/2^-1/2$ eigenphase shift for $N_{\rm max}=4$ shows also the $7/2^-1/2_2$ resonance.}
\label{7Liconv5-7-}
\end{figure}

We now turn to the description of the continuum part of the $^7$Li spectrum. We find that the convergence of the eigenphase shifts of the lowest two resonances ($7/2^-1/2_1$ and $5/2^-1/2_1$) with respect to the size of the model space ($N_\mathrm{max}$) is quite reasonable (Fig.~\ref{7Liconv5-7-}). For the eigenphase shifts of the other negative-parity resonances the convergence with respect to $N_\mathrm{max}$ is less satisfactory (Fig.~\ref{7Liconv5-7-}). Unlike the bound-state energies, the convergence of continuum properties does not follow an exponential pattern, making it difficult to extrapolate.

For the eigenphase shifts of the positive-parity resonances the convergence with respect to $N_{\rm max}$ is more challenging. For $N_{\rm max}=4(5)$ (the value in parentheses was used to calculate the NCSM eigenstates of $^7$Li) these states were obtained as bound states. The $1/2^+1/2$ state was obtained as a bound state also for $N_{\rm max}=6(7)$.

\begin{figure}
\includegraphics[width=\linewidth]{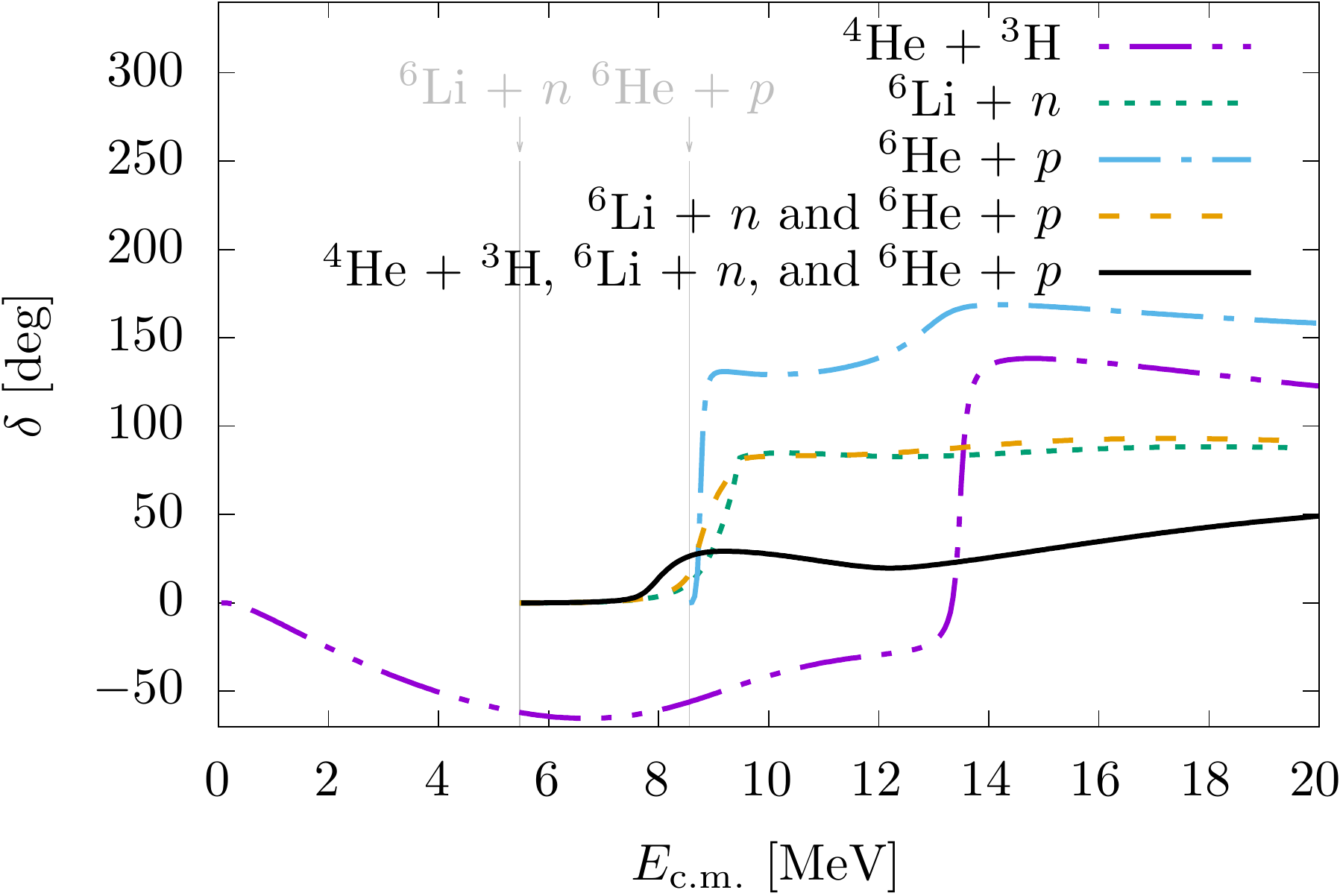}
\caption{The eigenphase shifts of the $1/2^+1/2$ resonance in $^7$Li as functions of the energy in the center-of-mass frame relative to the $^4$He + $^3$H threshold calculated for different choices of partitions. The vertical lines denote the calculated $^6$Li + $n$ and $^6$He + $p$ thresholds.}
\label{7Liconv1+}
\end{figure}

The effect of coupling partitions on the eigenphase shift of the $1/2^+1/2$ resonance is shown in Fig.~\ref{7Liconv1+}. For all the choices of partitions we obtain a resonance close to the $^6$He + $p$ threshold, nearby which the first $1/2^+1/2$ NCSM eigenstate appears. In particular, in the $^4$He + $^3$H calculation we obtain a broad resonance, which becomes narrower after coupling with the partitions $^6$Li + $n$ and $^6$He + $p$ due to the appearance of the $^6$He + $p$ threshold. The $^4$He + $^3$H and $^6$He + $p$ calculations also predict a second resonance which is due to coupling to the second $1/2^+1/2$ NCSM eigenstate.

\begin{figure}
\includegraphics[width=\linewidth]{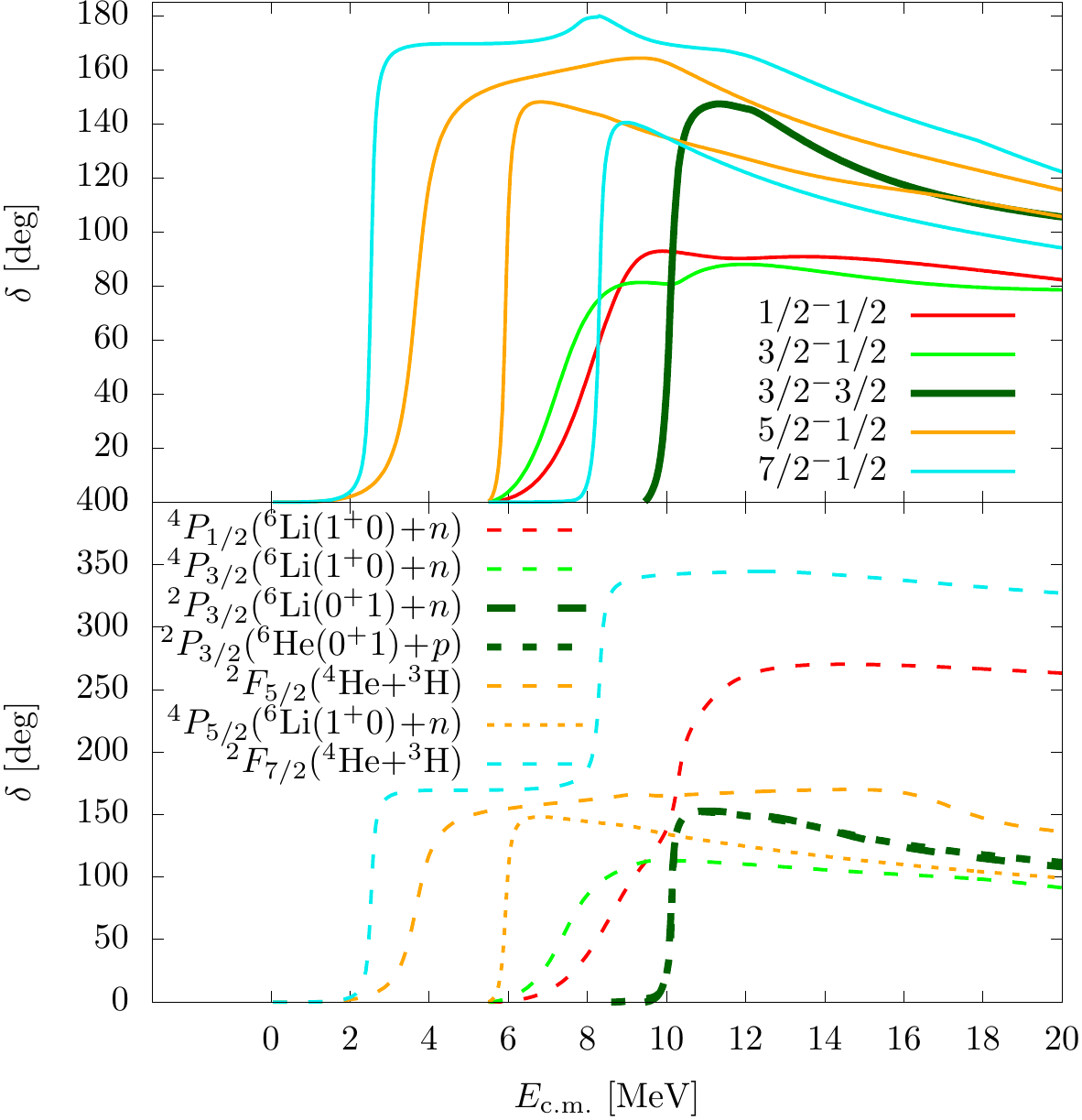}
\caption{The calculated eigenphase shifts and diagonal phase shifts corresponding to the experimentally observed resonances in $^7$Li as functions of the energy in the center-of-mass frame relative to the $^4$He + $^3$H threshold. The upper panel shows the eigenphase shifts for different values of the total angular momentum, parity, and isospin $J^{\pi}T$, and the lower panel shows the diagonal phase shifts for different channels.}
\label{fig1}
\end{figure}

The calculated eigenphase shifts and diagonal phase shifts of the experimentally observed resonances in $^7$Li, namely the $7/2^-1/2_1$, $5/2^-1/2_1$, $5/2^{-}1/2_{2}$, $3/2^{-}1/2_{2}$, $1/2^{-}1/2_{2}$, $7/2^{-}1/2_{2}$, and $3/2^{-}3/2_{1}$ states, are shown in Fig.~\ref{fig1}. The NCSMC calculation yields all the experimentally observed resonances in the correct order.

The centroids, $E_r$, of the resonances relative to the $^4$He + $^3$H threshold and their widths, $\Gamma$, were obtained by finding poles of the scattering matrix analytically continued to complex energy plane. For the experimentally observed resonances, we report their widths in Table~\ref{tab2}. The spectrum calculated within the NCSMC by coupling the three partitions is closer to the experimental data than the spectrum calculated with the $^4$He + $^3$H partition only (see Fig.~\ref{fig4}). The calculated widths are consistent with the experimental values, namely, the narrow (broad) resonances are reproduced as narrow (broad) (see Table~\ref{tab2}).

\begin{table}
\caption{Widths $\Gamma$ (in MeV) of the experimentally observed resonances in $^{7}$Li. The results of the NCSMC calculation coupling the three partitions are shown along with the results of the calculations taking into account the partitions separately. Energies $E_r$ relative to the $^4$He + $^3$H threshold (in MeV) of the resonances obtained in the coupled-partition calculation are shown as well. Experimental data are from Ref.~\cite{Tilley2002}.}
\label{tab2}
\begin{tabularx}{\columnwidth}{c c c c *4{>{\centering\arraybackslash}X}}
\hline
\hline
  & $^4$He+$^3$H & $^6$He+$p$ & $^6$Li+$n$ & \multicolumn{2}{c}{coupled} & \multicolumn{2}{c}{Exp.} \\
 $J^{\pi}T_i$ & $\Gamma$ & $\Gamma$ & $\Gamma$ & $E_r$ & $\Gamma$ & $E_r$ & $\Gamma$ \\ \hline
 $7/2^-1/2_1$ & 0.2 &  &  & 2.5 & 0.2 & 2.19 & 0.07 \\
 $5/2^-1/2_1$ & 0.8 &  &  & 3.7 & 0.7 & 4.14 & 0.92 \\
 $5/2^-1/2_2$ & 0.006 &  & 0.2 & 5.9 & 0.2 & 4.99 & 0.08 \\
 $3/2^-1/2_2$ & 0.3 &  & 1.7 & 7.2 & 2.1 & 6.28 & 4.71 \\
 $1/2^-1/2_2$ & not found &  & 2.6 & 8.1 & 2.4 & 6.62 & 2.75 \\
 $7/2^-1/2_2$ & 0.4 &  & 0.04 & 8.3 & 0.2 & 7.10 & 0.44 \\
 $3/2^-3/2_1$ &  & 0.4 & 0.4 & 10.1 & 0.3 & 8.77 & 0.26 \\
 \hline
 \hline
\end{tabularx}
\end{table}

In Fig.~\ref{fig4} we can see that the energies of the resonances present a stronger dependence on the choice of partition included in the calculation than the bound-state energies. In particular, the energies obtained from the $^6$He + $p$ calculation differ more significantly from those obtained from the $^6$Li + $n$ calculation. They are farther from the experimental data, except for the $5/2^-1/2_1$ resonance, which lies below the $^6$Li + $n$ threshold, and the $3/2^-3/2_1$ resonance, which lies above the $^6$He + $p$ threshold. The $^6$Li + $n$ partition is expected to play a more significant role than the $^6$He + $p$ partition in describing resonances that lie above the $^6$Li + $n$ threshold and below the $^6$He + $p$ threshold, because these resonances can decay by $^6$Li + $n$ breakup--a process which is accounted for by the $^6$Li + $n$ partition. The calculated properties of the reproduced resonances are affected by the coupling of the partitions.

\begin{figure}
\includegraphics[width=\linewidth]{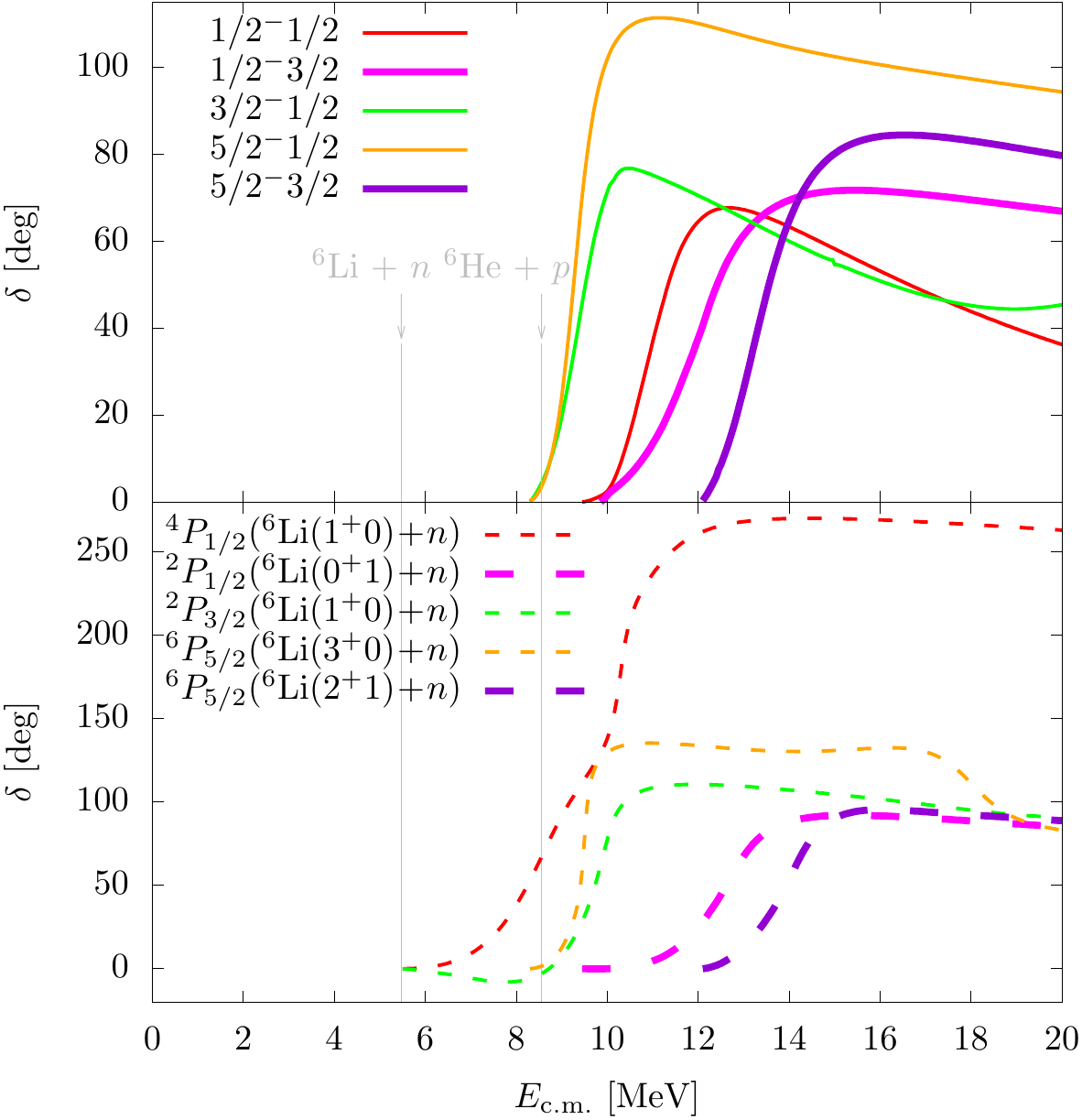}
\caption{The eigenphase shifts and diagonal phase shifts predicting new negative-parity resonances in $^7$Li as functions of the energy in the center-of-mass frame relative to the $^4$He + $^3$H threshold. The upper panel shows the eigenphase shifts for different values of the total angular momentum, parity, and isospin $J^{\pi}T$, and the lower panel shows the diagonal phase shifts for different channels. The vertical lines denote the calculated $^6$Li + $n$ and $^6$He + $p$ thresholds.}
\label{fig2}
\end{figure}

\begin{figure}
\includegraphics[width=\linewidth]{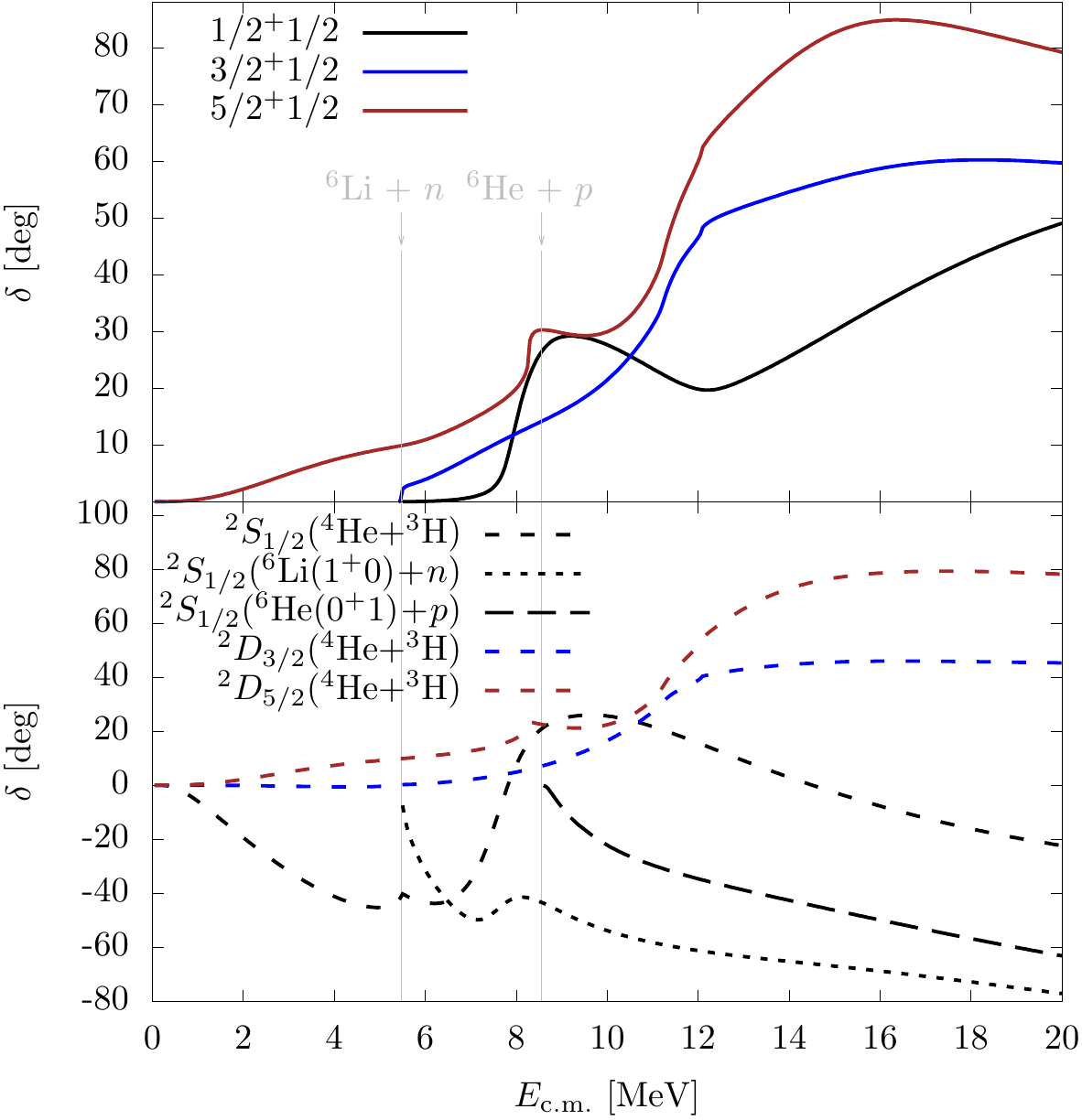}
\caption{Eigenphase shifts and diagonal phase shifts predicting new positive-parity resonances in $^7$Li as functions of the energy in the center-of-mass frame relative to the $^4$He + $^3$H threshold. The upper panel shows the eigenphase shifts with resonant behavior for different values of the total angular momentum, parity, and isospin $J^{\pi}T$, and the lower panel shows the diagonal phase shifts for different channels. The vertical lines denote the calculated $^6$Li + $n$ and $^6$He + $p$ thresholds.}
\label{fig3}
\end{figure}

The remaining calculated eigenphase shifts and diagonal phase shifts that have resonant behavior and thus predict new resonances in $^7$Li that have not been experimentally observed are shown in Fig.~\ref{fig2} and Fig.~\ref{fig3} for the negative and positive parity, respectively. In the energy region up to 20 MeV above the $^4$He + $^3$H threshold five resonances with negative parity and $J^{\pi}T=1/2^-1/2$, $1/2^-3/2$, $3/2^-1/2$, $5/2^-1/2$, $5/2^-3/2$ are predicted (Fig.~\ref{fig2}) along with three positive-parity resonances with $J=1/2$, $3/2$, $5/2$ and $T=1/2$ (Fig.~\ref{fig3}). From the diagonal phase shifts we can see that the negative-parity resonances appear in $P$ waves in $^6$Li + $n$ channels and the positive-parity resonances are dominated by $S$ and $D$ waves in the $^4$He + $^3$H channel.

As a summary, the excitation energies of the predicted resonances are shown in Fig.~\ref{fig4} and their widths are shown in Table~\ref{tab3}. The results of the NCSMC calculations including the three partitions separately are also shown. The properties of these resonances are affected by the coupling of the partitions.

\begin{table}
\caption{Widths $\Gamma$ (in MeV) of the predicted resonances in $^{7}$Li calculated within the NCSMC. Energies $E_r$ relative to the $^4$He + $^3$H threshold (in MeV) of the resonances obtained in the coupled-partition calculation are shown as well.}
\label{tab3}
\begin{tabularx}{\columnwidth}{*6{>{\centering\arraybackslash}X}}
 \hline
 \hline
 & $^4$He+$^3$H & $^6$He+$p$ & $^6$Li+$n$ & \multicolumn{2}{c}{coupled}  \\
$J^{\pi}T_i$ & $\Gamma$ & $\Gamma$ & $\Gamma$ & $E_r$ & $\Gamma$ \\
\hline
 $1/2^+1/2_1$ & 5.6 & 0.1       & 1.2 & 6.9  & 4.9 \\
 $5/2^-1/2_3$ & 0.2 & - & 1.0 & 9.3  & 0.9 \\
 $3/2^-1/2_3$ & 0.3 & 0.5       & 1.7 & 9.4  & 2.2 \\
 $5/2^+1/2_1$ & 5.1 & 0.5       & 1.2 & 9.7  & 6.3 \\
 $1/2^-1/2_3$ & 0.3 & 0.1       & 1.5 & 10.8 & 2.1 \\
 $3/2^+1/2_1$ & 0.1 & 1.1       & 0.1 & 11.2 & 5.0 \\
 $1/2^-3/2_1$ &     & 2.4       & 2.8 & 11.8 & 3.5 \\
 $5/2^-3/2_1$ &     & 2.1       & 2.7 & 13.2 & 2.2 \\
 \hline
 \hline
\end{tabularx}
\end{table}

To conclude this section, we address the discrepancy between a NCSMC prediction and recent experimental data pointed out in Ref.~\cite{Dronchi2023}. In Ref.~\cite{Vorabbi2019} Vorabbi \textit{et al.} carried out calculations for $^7$Li within the NCSMC including the three partitions in separate calculations. They predicted a new resonance just above the $^6$He + $p$ threshold in the $^{2}S_{1/2}$ partial wave in the $^6$Li + $n$ and $^6$He + $p$ channels built on the ground state of the target. However, in Ref.~\cite{Dronchi2023} no $S$-wave $1/2^+$ resonance was experimentally observed in the $^6$He + $p$ channel. Our calculation coupling the three partitions also predicts a $1/2^+1/2$ resonance as shown in Fig.~\ref{fig3}. However, it is below the $^6$He + $p$ threshold and dominated by the $^2S_{1/2}$ partial wave in the $^4$He + $^3$H channel. This may explain why no $S$-wave $1/2^+$ resonance was experimentally observed in the $^6$He + $p$ channel.

We did not obtain the $1/2^+$ resonance when using only the binary-cluster RGM basis states, which indicates that the resonance comes from the NCSM description of $^7$Li. This is in accordance with Ref.~\cite{Arai2001}, which studies $^6$He + $p$ reactions in a microscopic multicluster model and predicts no $1/2^+$ resonance in $^7$Li.

\subsubsection{Cross sections}
\label{sec:cross}

\begin{figure}
\includegraphics[width=\linewidth]{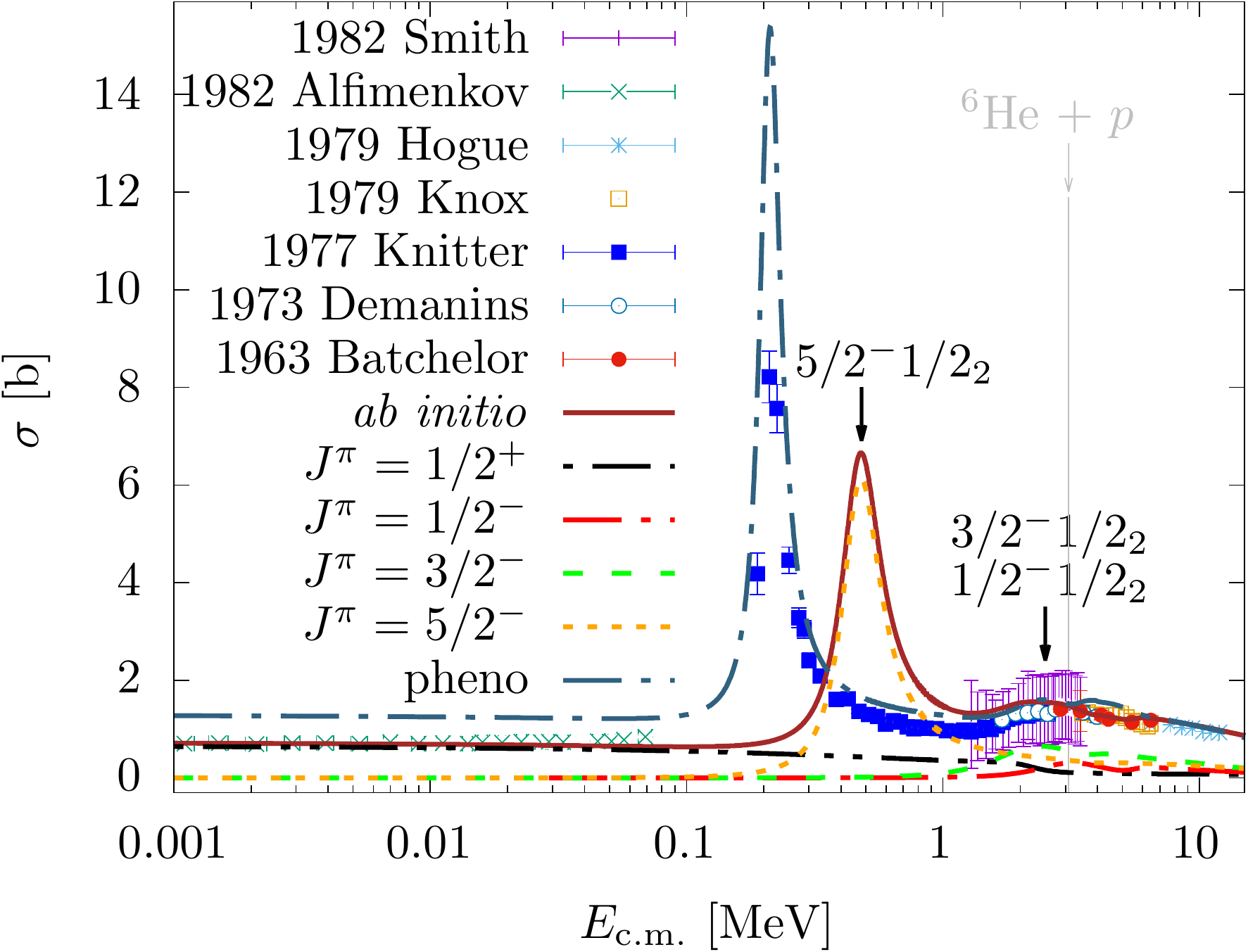}
\caption{The cross section of the elastic scattering of neutrons on $^6$Li as a function of the kinetic energy in the center-of-mass frame calculated within the NCSMC (solid line) along with experimental data~\cite{Smith1982,Alfimenkov1982,Hogue1979,Knox1979,Knitter1977,Demanins1973,Batchelor1963} and results calculated by restricting the total angular momentum and parity $J^{\pi}$ to $1/2^+$, $1/2^-$, $3/2^-$, and $5/2^-$ (dashed lines). Resonances corresponding to the peaks in the calculated cross section are indicated. The vertical gray line denotes the calculated $^6$He + $p$ threshold. The cross section calculated phenomenologically adjusting NCSM energies is shown as well and denoted by ``pheno".}
\label{sigmael}
\end{figure}

Fig.~\ref{sigmael} shows the calculated cross section of the elastic scattering of neutrons on $^6$Li as a function of the kinetic energy in the center-of-mass frame (solid line) along with experimental data. There are two visible peaks, with the first corresponding to the $5/2^-1/2_2$, and the second to the $3/2^-1/2_2$ resonances with a small contribution from the $1/2^-1/2_2$ resonance. The two-peak structure is reflected in the experimental data, albeit at slightly different energies, reflecting the inaccuracies of the interaction chosen for this work. To better describe the cross section, in Sect.~\ref{sec:pheno} we consider phenomenological adjustments of the Hamiltonian kernel appearing in the NCSMC equations.

\begin{figure}
\includegraphics[width=\linewidth]{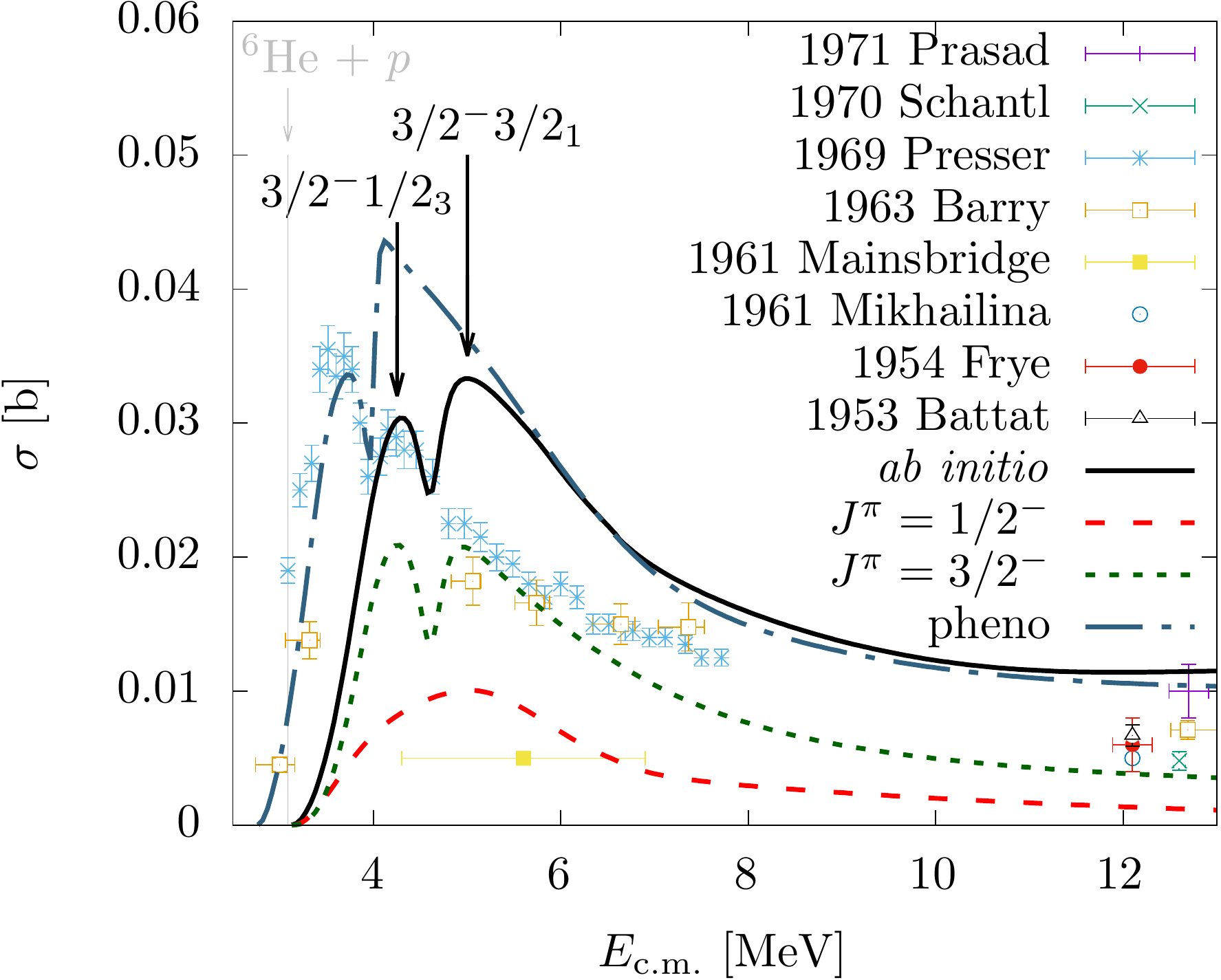}
\caption{The cross section of the $^6$Li($n,p)^6$He reaction as a function of the kinetic energy in the center-of-mass-frame calculated within the NCSMC (solid line) along with experimental data~\cite{Prasad1971,Schantl1970,Presser1969,Barry1963,Mainsbridge1961,Mikhailina1961,Frye1954,Battat1953} and results calculated by restricting the total angular momentum and parity $J^{\pi}$ to $1/2^-$ and $3/2^-$ (dashed lines). Resonances corresponding to the peaks in the calculated cross section are indicated. The vertical gray line denotes the calculated $^6$He + $p$ threshold. The cross section calculated phenomenologically adjusting NCSM energies is shown as well and denoted by ``pheno".}
\label{fig5}
\end{figure}

An advantage of the coupled-partition calculation is the ability to calculate the cross section of the charge-exchange reaction $^6$Li($n,p)^6$He. This cross section (for the reaction going from the ground state of $^6$Li to the ground state of $^6$He) is shown in Fig.~\ref{fig5} as a function of the kinetic energy in the center-of-mass frame along with experimental data. The calculation reproduces the reaction threshold (corresponding to the difference 3.1 MeV between the neutron and proton separation energies depicted in Fig.~\ref{fig4}, which is in good agreement with the experimental value 2.7255 MeV~\cite{Tilley2002}) and the overall shape of the cross section. The calculated cross section has a two-peak structure arising from the (predicted) $3/2^-1/2_3$ and the (reproduced) $3/2^-3/2_1$ resonances with a contribution from the (predicted) $1/2^-1/2_3$ resonance. This structure is mirrored in the Presser data~\cite{Presser1969} and the  assignment of the resonance angular momenta, parities, and isospins agrees with that of Ref.~\cite{Presser1969}. However, again, the calculated cross section peaks appear at higher energy compared to the measured cross section pointing to a systematic deviation across the spectrum. Again, we will attempt to correct this shape in Sect.~\ref{sec:pheno} via phenomenological adjustments of the Hamiltonian kernel appearing in the NCSMC equations.

\begin{figure}
\includegraphics[width=\linewidth]{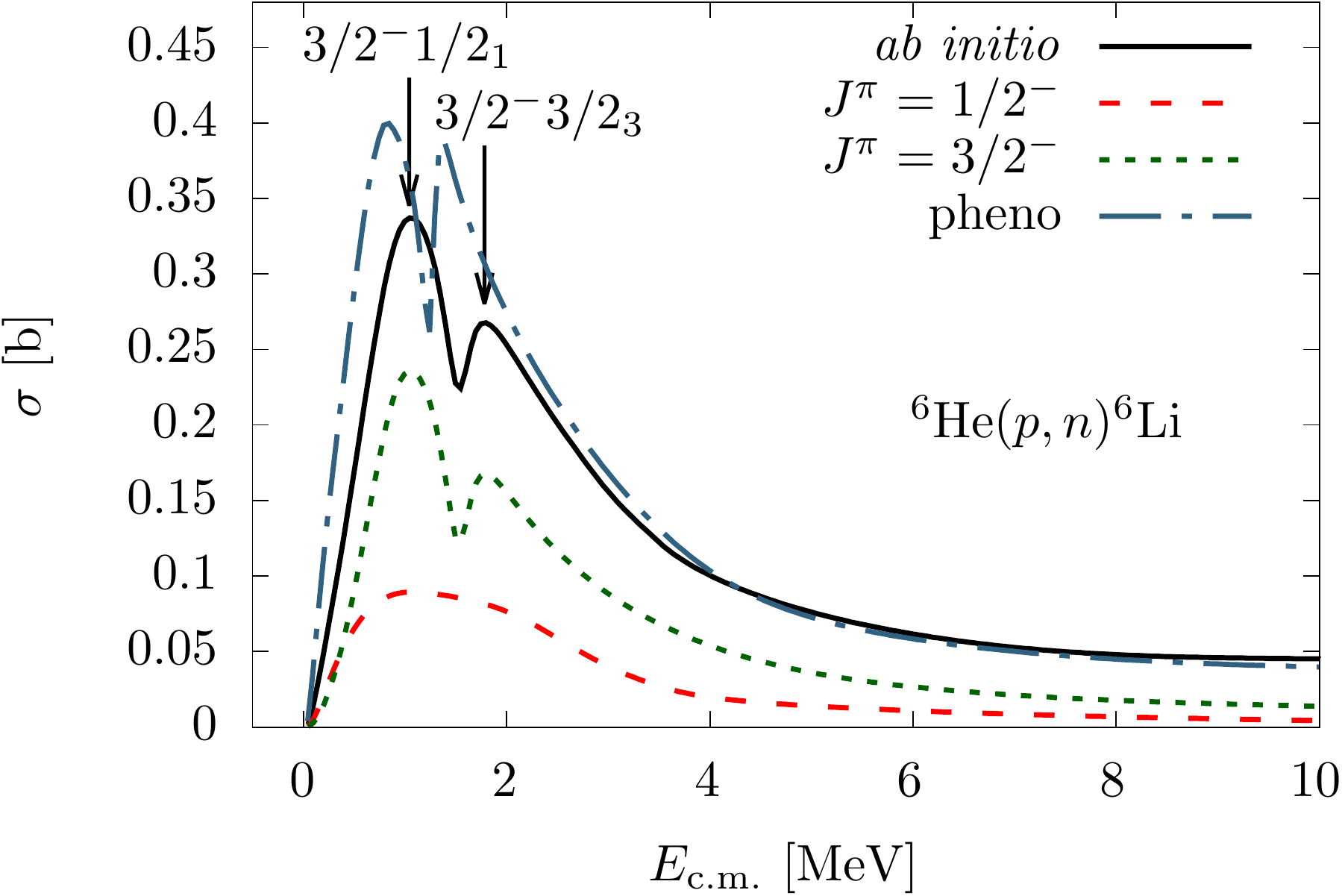}
\caption{Calculated cross section of the $^6$He($p,n)^6$Li reaction as a function of the kinetic energy in the center-of-mass frame (solid line) along with results calculated by restricting the total angular momentum and parity $J^{\pi}$ to $1/2^-$ and $3/2^-$ (dashed lines). Resonances corresponding to the peaks are indicated. The cross section calculated phenomenologically adjusting NCSM energies is shown as well and denoted by ``pheno".}
\label{fig:pn}
\end{figure}

\begin{figure}
\includegraphics[width=\linewidth]{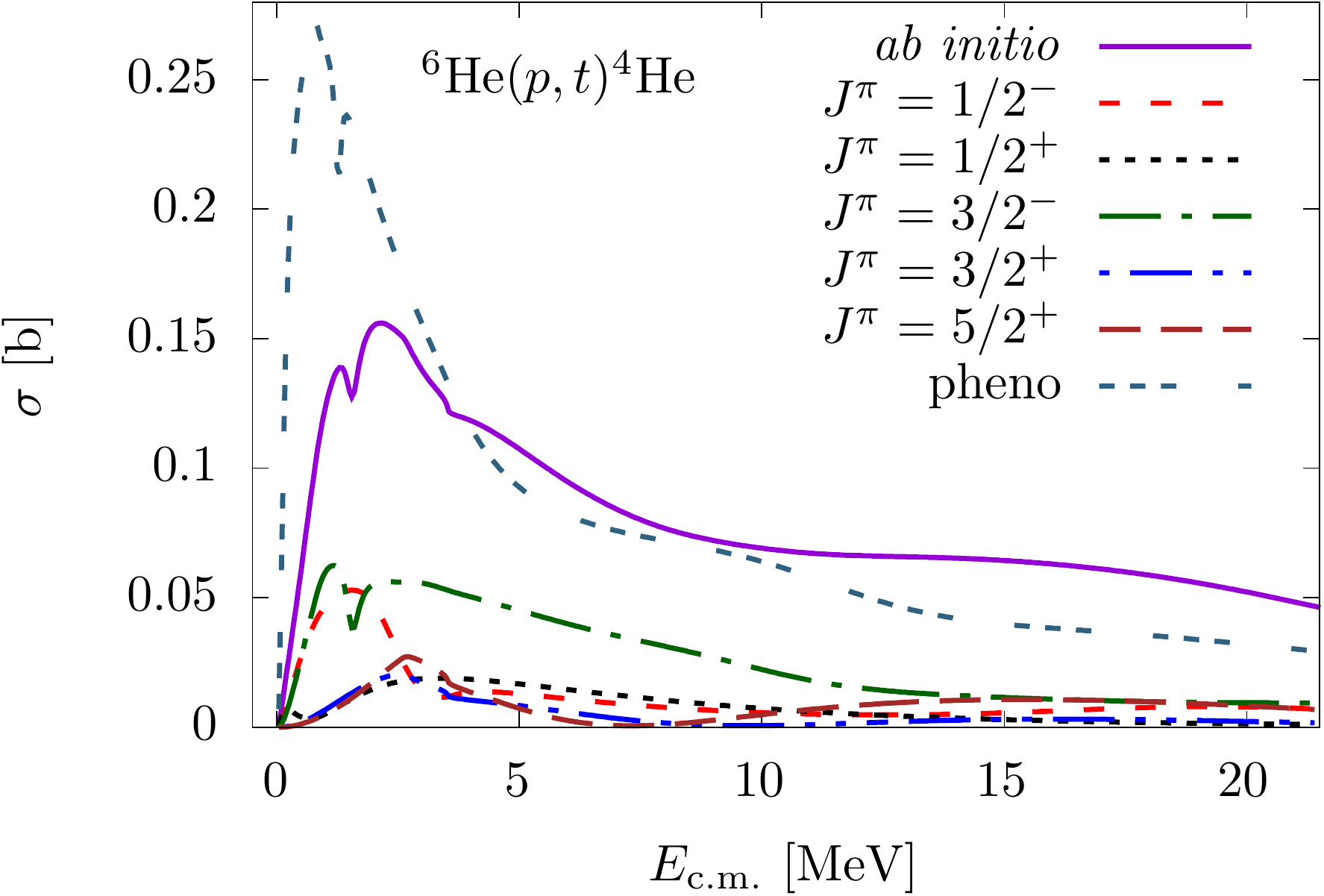}
\caption{Calculated cross section of the $^6$He($p,t)^4$He reaction as a function of the kinetic energy in the center-of-mass frame (solid line) along with results calculated by restricting the total angular momentum and parity $J^{\pi}$ to specific values (dashed lines). The cross section calculated phenomenologically adjusting NCSM energies is shown as well and denoted by ``pheno".}
\label{fig:pt}
\end{figure}

To provide theoretical predictions for a comparison with future measurement of $^6$He + $p$ reactions planned at TRIUMF, Figs.~\ref{fig:pn} and~\ref{fig:pt} show the calculated cross sections of the reactions $^6$He($p,n)^6$Li and $^6$He($p,t)^4$He (solid lines), calculation of which is allowed by the coupling of the three partitions. Similarly to the $^6$Li($n,p)^6$He reaction, the cross sections have two peaks corresponding to the $3/2^-1/2_3$ and $3/2^-3/2_1$ resonances with contributions from other $J^{\pi}$ channels, as indicated by results calculated by restricting $J^{\pi}$ to given values (dashed lines in Figs.~\ref{fig:pn} and~\ref{fig:pt}). The kink in the ($p,t$) cross section in Fig.~\ref{fig:pt} at energy $E_{\rm c.m.}\approx3.5$ MeV appears due to opening of the $^6$Li($2^+1)+n$ channel.

\begin{figure}
\includegraphics[width=\linewidth]{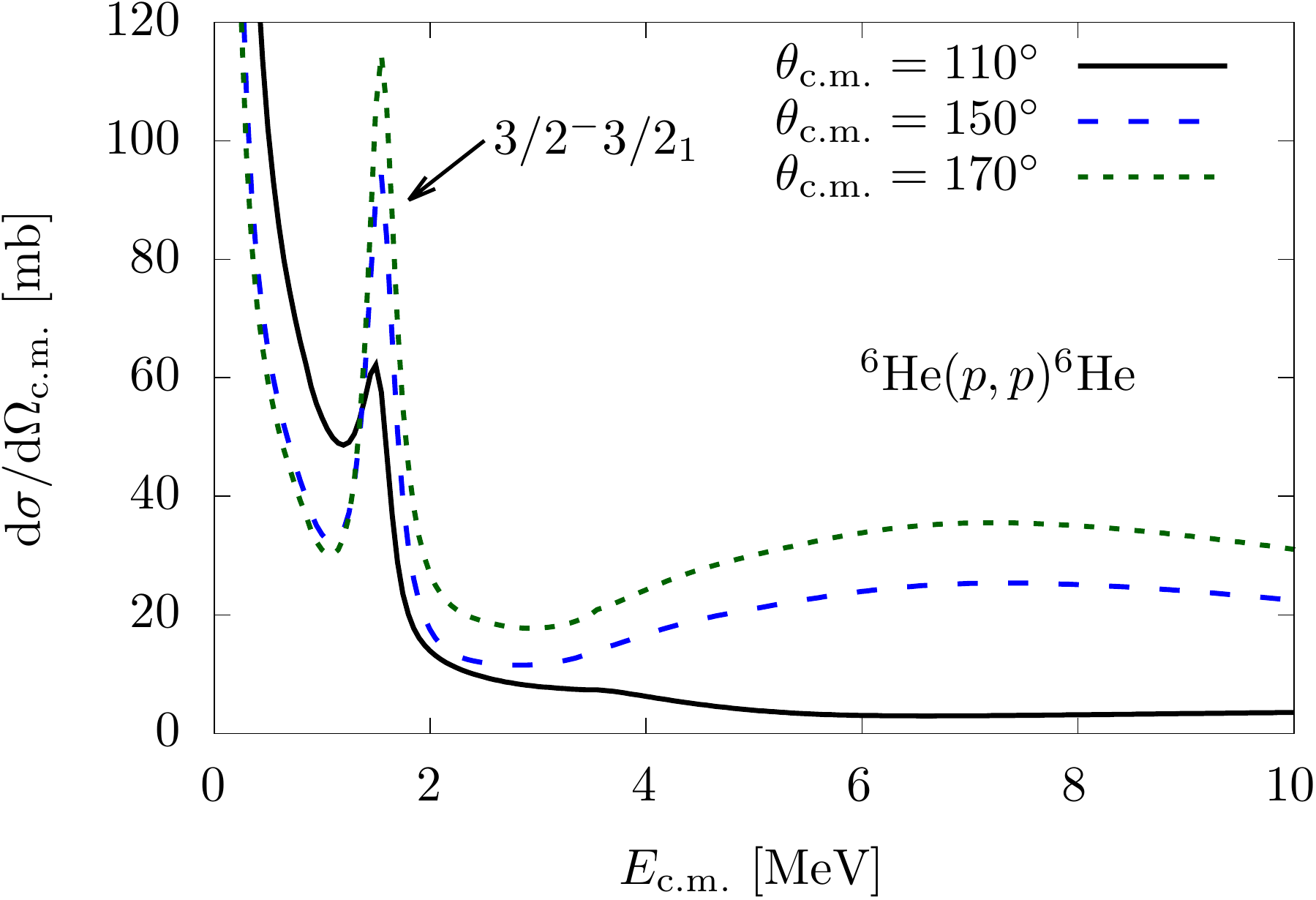}
\caption{Calculated differential cross sections of the elastic scattering of protons on $^6$He as functions of the kinetic energy in the center-of-mass frame for three different scattering angles. The peaks correspond to the $3/2^-3/2_1$ resonance.}
\label{fig:ds}
\end{figure}

\begin{figure}
\includegraphics[width=\linewidth]{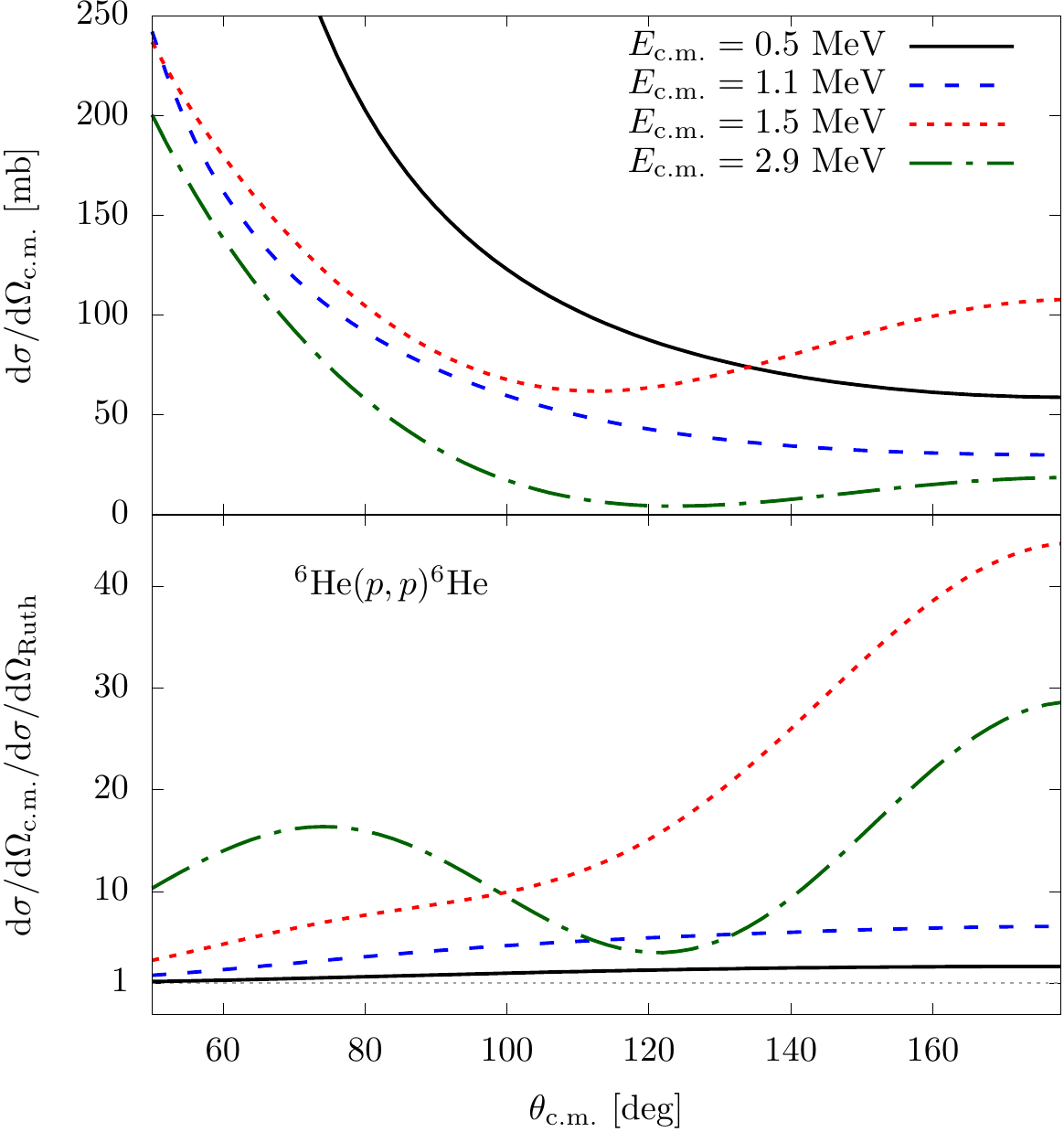}
\caption{Calculated differential cross sections of the elastic scattering of protons on $^6$He as functions of the scattering angle for different values of the kinetic energy in the center-of-mass frame $E_{\rm c.m.}$ (top) and ratios of the cross sections to the Rutherford cross section (bottom).}
\label{fig:dsdO}
\end{figure}

To further support the upcoming experiment planned at TRIUMF, Fig.~\ref{fig:ds} shows the calculated differential cross section of the elastic scattering of protons on $^6$He as a function of the kinetic energy in the center-of-mass frame for backward scattering angles $\theta_{\rm c.m.}=110^{\circ},\,150^{\circ},\,170^{\circ}$, and Fig.~\ref{fig:dsdO} shows the calculated differential cross section of the scattering as a function of the scattering angle for different values of the kinetic energy in the center-of-mass frame along with ratios of the cross sections to the Rutherford cross section. In Fig.~\ref{fig:ds} we can see that for low energies the differential cross sections increase with decreasing energy in accordance with the Rutherford formula~\cite{thompsonnunes}. The differential cross sections have a peak, which corresponds to the $3/2^-3/2_1$ resonance and is much stronger for $\theta_{\rm c.m.}=150^{\circ},\,170^{\circ}$ than for $\theta_{\rm c.m.}=110^{\circ}$. This can be explained in the following way. The resonance is a $P$-wave resonance (see lower panel of Fig.~\ref{fig1}) and therefore suppressed for $\theta_{\rm c.m.}=110^{\circ}$ by the Legendre polynomial $P_1(\cos\theta_{\rm c.m.})=\cos\theta_{\rm c.m.}$. For $\theta_{\rm c.m.}=150^{\circ}$ the Legendre polynomial is greater and therefore the resonance is stronger. For $\theta_{\rm c.m.}=170^{\circ}$ the Legendre polynomial is even greater and therefore the resonance is even stronger.

The upper panel of Fig.~\ref{fig:dsdO} contains differential cross sections for four energies around the peak in Fig.~\ref{fig:ds}. We can see that for low scattering angles the differential cross sections increase with decreasing scattering angle in accordance with the Rutherford formula, and for a given scattering angle they oscillate with increasing energy in accordance with Fig.~\ref{fig:ds}. The lower panel of Fig.~\ref{fig:dsdO} shows the ratios of the calculated differential cross sections to the Rutherford cross section. For the lowest energy, the calculated differential cross section is close to the Rutherford cross section (the ratio is close to 1). With increasing energy and scattering angle the discrepancy from the Rutherford cross section (the ratio) increases.

\begin{figure}
\includegraphics[width=\linewidth]{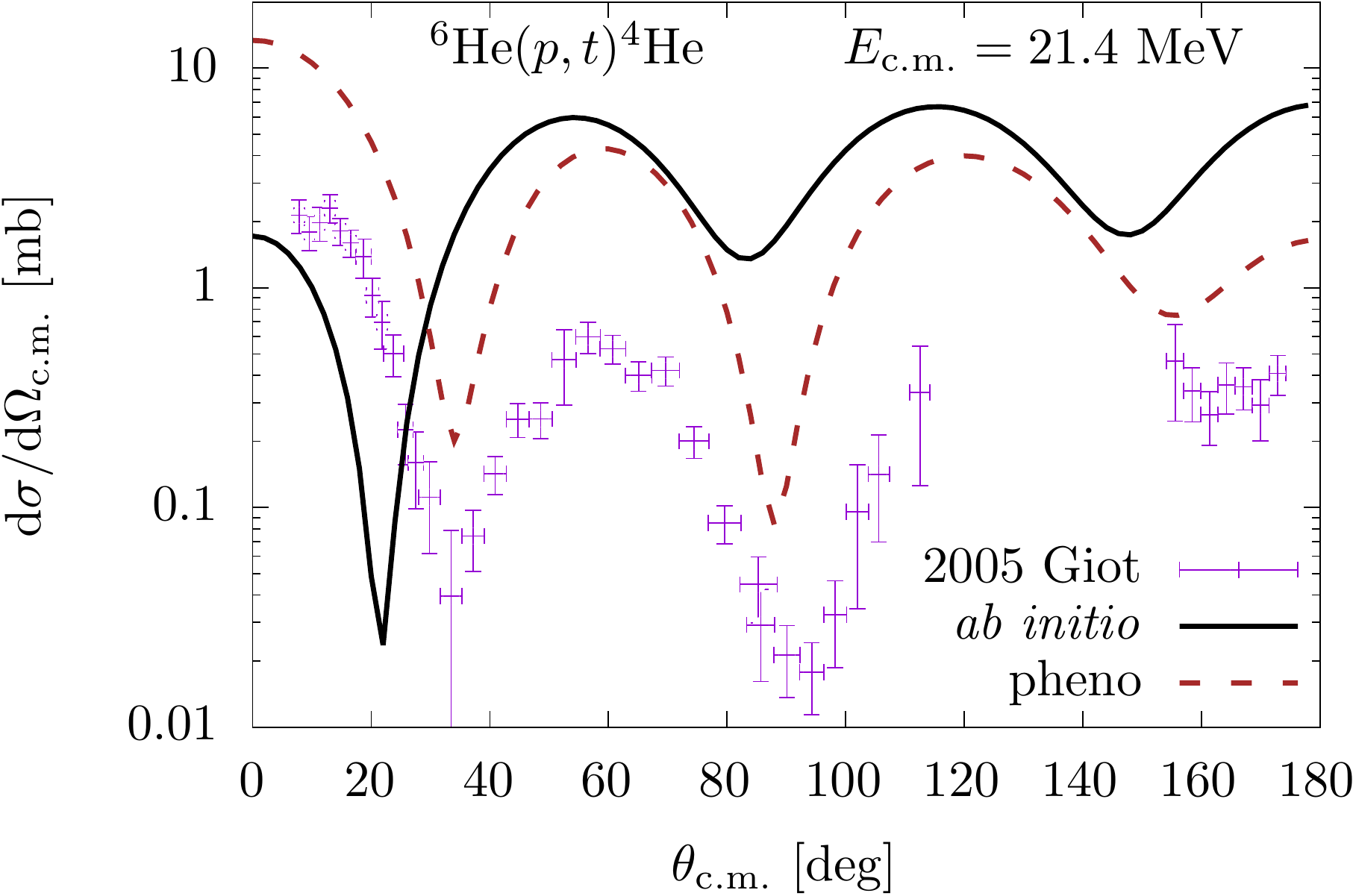}
\caption{Differential cross section of the $^6$He($p,t)^4$He reaction as a function of the scattering angle for kinetic energy in the center-of-mass frame $E_{\rm c.m.}=21.4$ MeV calculated \textit{ab initio} (solid line) and phenomenologically adjusting NCSM energies (dashed line) along with experimental data~\cite{Giot2005}.}
\label{fig:dspt}
\end{figure}

Fig.~\ref{fig:dspt} shows the calculated differential cross section of the $^6$He($p,t)^4$He reaction as a function of the scattering angle for kinetic energy in the center-of-mass frame $E_{\rm c.m.}=21.4$ MeV (solid line) along with experimental data~\cite{Giot2005}. The calculated cross section reproduces the overall shape of the measured cross section, however, it does not reproduce the positions of the minima and maxima. To better describe the cross section we consider phenomenological adjustments of the Hamiltonian kernel in the next section.

\subsection{Phenomenological adjustments}
\label{sec:pheno}

The NCSMC equations are expressed via integral kernels, namely the norm and Hamiltonian kernels~\cite{Baroni2013C,Navratil2016}, with the Hamiltonian kernel being a function of the energies of the targets and the aggreagate-nucleus eigenstates calculated within the NCSM. Here we consider phenomenological adjustments of these NCSM energies in the Hamiltonian kernel to improve the description of the cross sections presented in Sect.~\ref{sec:cross}. A similar approach was used in Refs.~\cite{Eraly2016,Hupin2015,Hupin2019,Hebborn2022,Kravvaris2023,Atkinson2025}.

First we made adjustments of the NCSM energies of the ground and excited states of $^6$Li and $^6$He to reproduce the experimentally observed spectra and the corresponding $^6$Li + $n$ and $^6$He + $p$ decay thresholds with respect to the $^4$He + $^3$H threshold. 

Next, we adjusted the NCSM energies of $^7$Li states. In particular, we adjusted the energies of the states $5/2^-1/2_2$ and $3/2^-3/2_1$ to reproduce their distances to the $^6$Li + $n$ threshold. In this way we reproduced the positions of the peak in the elastic neutron scattering cross section (Fig.~\ref{sigmael}) and the second peak in the $^6$Li($n,p)^6$He cross section (Fig.~\ref{fig5}). We also adjusted the energy of the state $3/2^-1/2_3$ to reproduce the position of the first peak in the $^6$Li($n,p)^6$He cross section and the energies of the states $3/2^-1/2_2$ and $1/2^-1/2_2$ to improve the agreement with the experimental elastic neutron scattering cross section at high energy and of the state $1/2^-1/2_3$ to better describe the overall shape of the $^6$Li($n,p)^6$He cross section.

The resulting cross sections are shown in Figs.~\ref{sigmael} and~\ref{fig5} for the elastic neutron scattering and the $^6$Li($n,p)^6$He reaction, respectively, and denoted by ``pheno". In Fig.~\ref{sigmael} we can see that the new elastic cross section presents a peak at the same energy as the experimental cross section and agrees fairly well with the high-energy data. However, the height of the peak and values of the cross section at low energy overestimate the experimental values. Given the existence of a sub-threshold 5/2$^-$ resonance it is possible that the exact structure of the states is not well-reproduced with the current nuclear interaction, leading to a much larger neutron component and thus a significantly larger cross section at resonance. In Fig.~\ref{fig5} we can see that the new charge-exchange cross section reproduces the position of the two peaks, however, it also overestimates the height of the second peak. Again, a similar reason could be behind this discrepancy; future calculations with multiple interactions will provide the needed clarity. In Fig.~\ref{sigmael} we can see that the new elastic cross section overestimates the experimental values at low energy. We tried to fix this by shifting up the energy of the NCSM state corresponding to the $1/2^+1/2$ resonance, however, it had no effect on the cross section.

The cross sections of the reactions $^6$He($p,n)^6$Li and $^6$He($p,t)^4$He obtained after the phenomenological adjustments are shown in Figs.~\ref{fig:pn},~\ref{fig:pt}, and~\ref{fig:dspt} and are again denoted by ``pheno". In Fig.~\ref{fig:pn} we can see that the overall shape of the $^6$He($p,n)^6$Li cross section changed only slightly. In Fig.~\ref{fig:pt} we can see that the $^6$He($p,t)^4$He cross section is enhanced at low energy and preserves the two-peak shape. In Fig.~\ref{fig:dspt} we can see that the new differential cross section of the $^6$He($p,t)^4$He reaction better reproduces the positions of the minima and maxima. It overestimates the experimental values, but the energy $E_{\rm c.m.}=21.4$ MeV is beyond applicability of the current setup due to omitted channels which are open at this high energy. However, there are no experimental data at lower energy for comparison.

\begin{figure}
\includegraphics[width=\linewidth]{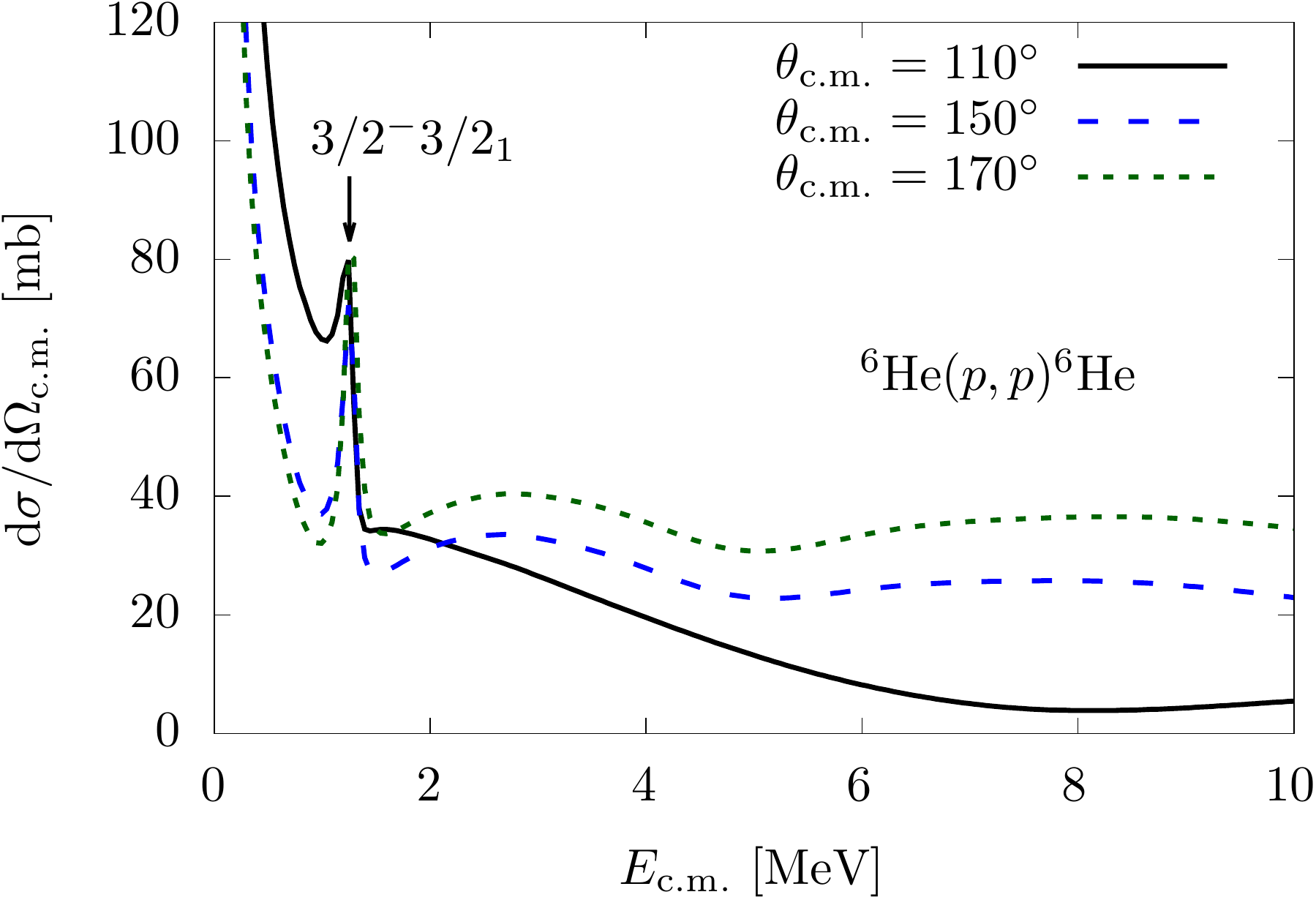}
\caption{Differential cross sections of the elastic scattering of protons on $^6$He as functions of the kinetic energy in the center-of-mass frame for three different scattering angles calculated phenomenologically adjusting NCSM energies. The peak corresponds to the $3/2^-3/2_1$ resonance.}
\label{figdspheno}
\end{figure}

\begin{figure}
\includegraphics[width=\linewidth]{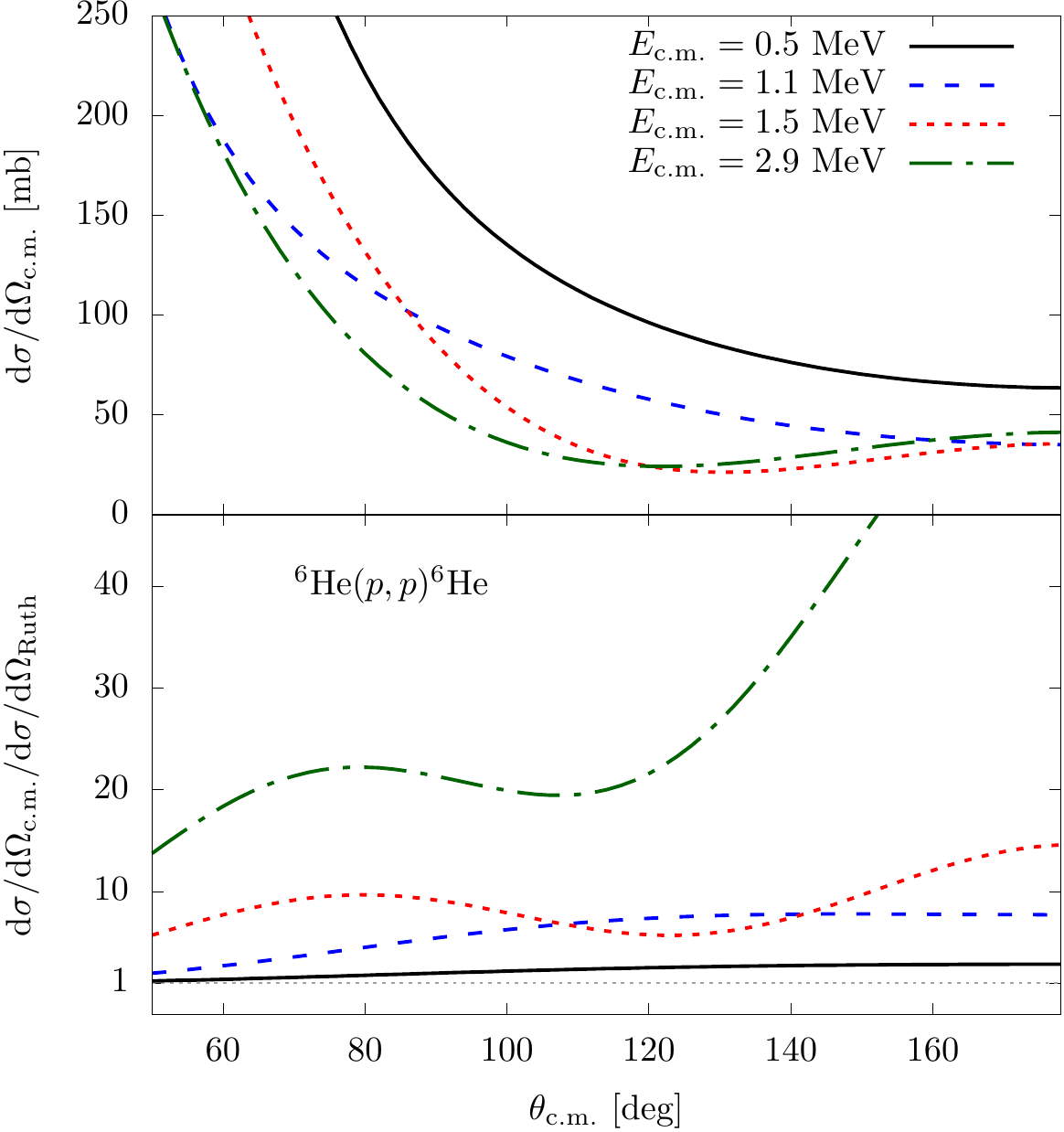}
\caption{Upper panel shows differential cross sections of the elastic scattering of protons on $^6$He as functions of the scattering angle for different kinetic energies in the center-of-mass frame calculated phenomenologically adjusting NCSM energies. The lower panel shows the ratios of the cross sections to the Rutherford cross section.}
\label{figdsdOpheno}
\end{figure}

The differential cross sections of the elastic scattering of protons on $^6$He obtained after the adjustments are shown in Fig.~\ref{figdspheno} (as functions of energy) and~\ref{figdsdOpheno} (as functions of the scattering angle $\theta_{\rm c.m.}$). The peaks in the differential cross sections as functions of energy (Fig.~\ref{figdspheno}) for $\theta_{\rm c.m.}=150^{\circ}, 170^{\circ}$ are less pronounced compared to the original result in Fig.~\ref{fig:ds}. Otherwise the differential cross sections have similar overall shape.

\begin{figure}
\includegraphics[width=\linewidth]{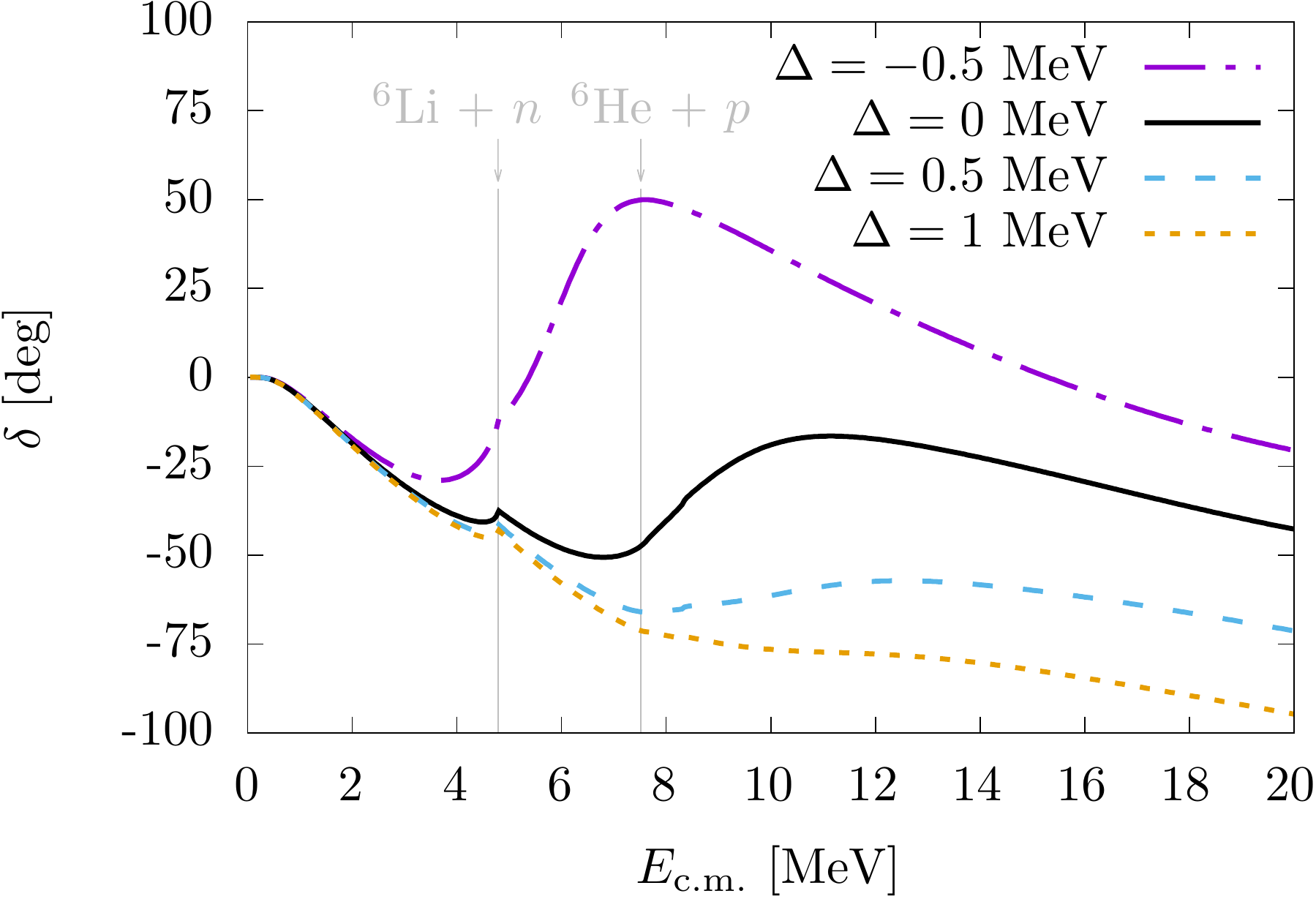}
\caption{Diagonal phase shifts corresponding to the $^2S_{1/2}$ partial wave in the $^4$He + $^3$H channel, where the $1/2^+1/2_1$ resonance appears, as functions of the energy in the center-of-mass frame relative to the $^4$He + $^3$H threshold calculated for the NCSM energy of the $1/2^+1/2_1$ state shifted by $\Delta$. The adjustments of the NCSM energies of the states of $^6$Li and $^6$He are made as well. The vertical lines denote the $^6$Li + $n$ and $^6$He + $p$ thresholds.}
\label{figps1+s}
\end{figure}

Fig.~\ref{figps1+s} shows the behavior of the phase shift corresponding to the $^2S_{1/2}$ partial wave in the $^4$He + $^3$H channel, where the $1/2^+1/2_1$ resonance appears, as we shift the NCSM energy of the corresponding state. Increasing the energy makes the resonance disappear. Decreasing the energy yields a narrower resonance at lower energy, however, this effect is shown only for illustration and is not anticipated to happen for $N_{\rm max}>11$. The phase shifts corresponding to the $^2D_{3/2}$ and $^2D_{5/2}$ partial waves in the $^4$He + $^3$H channel, where the $3/2^+1/2_1$ and $5/2^+1/2_1$ resonances appear, behave similarly as we shift the NCSM energies of the corresponding states.

\begin{figure}
\includegraphics[width=\linewidth]{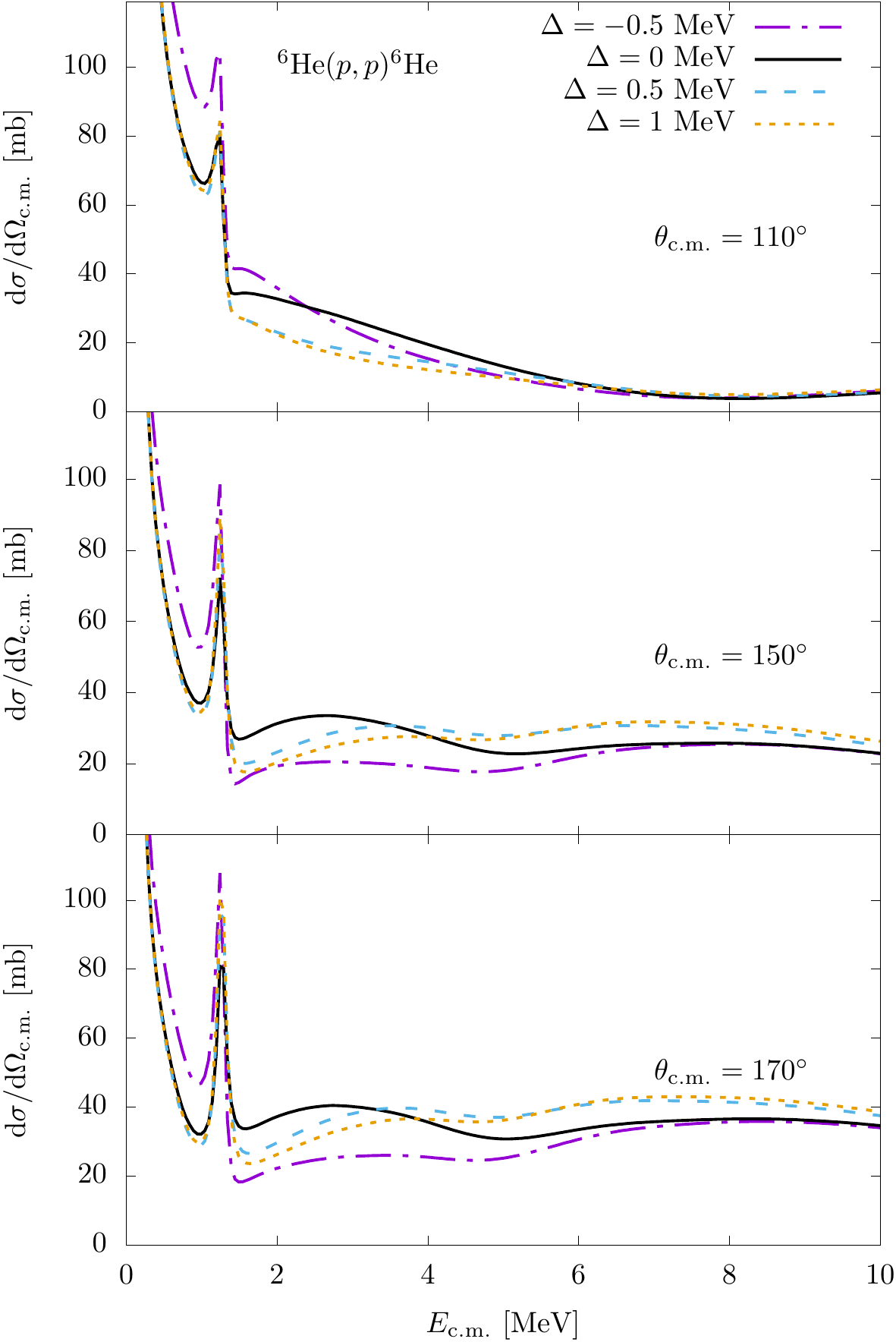}
\caption{Differential cross sections of the elastic scattering of protons on $^6$He as functions of the kinetic energy in the center-of-mass frame calculated for the NCSM energy of the $1/2^+1/2_1$ state shifted by $\Delta$. The NCSM energies of the other states are as in Fig.~\ref{figdspheno}.}
\label{figds1+s}
\end{figure}

The adjustments of the NCSM energy of the $1/2^+1/2$ state have an effect on the differential cross section of the elastic scattering of protons on $^6$He as a function of energy. This effect is shown in Fig.~\ref{figds1+s} for three different scattering angles. Decreasing the energy enhances the cross section at low energy including the $3/2^-3/2_1$ resonance, suppresses the cross section at energies above the resonance, and has no effect at high energies. Increasing the NCSM energy has no significant effect below the resonance, enhances the resonance, suppresses the cross section above the resonance, and enhances it at high energies (except $\theta_{\rm c.m.}=110^{\circ}$).

\section{Conclusions}
\label{sec4}

This work presents the first NCSMC coupled-channels calculation that includes three partitions. In particular, we carried out calculations for the $^7$Li system coupling the partitions $^4$He + $^3$H, $^6$Li + $n$, and $^6$He + $p$. This improves on the previous work of Ref.~\cite{Vorabbi2019}, where these same partitions were included in separate calculations. We find significant effects due to the partition coupling across structure and reaction observables; the centroids and widths of the resonances, the spectrum (Fig.~\ref{fig4}) and widths (Tables~\ref{tab2} and~\ref{tab3}) obtained in the coupled-partition calculation are different from the results obtained in the three single-partition calculations.

The previous NCSMC calculations for $^7$Li~\cite{Vorabbi2019} predicted an $S$-wave $1/2^+1/2$ resonance in the $^6$He + $p$ partition appearing just above the $^6$He + $p$ threshold, but the experimental search~\cite{Dronchi2023} for this resonance in the $^6$He + $p$ channel did not find it. Our calculation also predicts an $S$-wave $1/2^+1/2$ resonance in $^7$Li placing it, however,  below the $^6$He + $p$ threshold and dominated by the $^2S_{1/2}$ partial wave in the $^4$He + $^3$H channel. This can explain why no such resonance was experimentally found in the $^6$He + $p$ channel~\cite{Dronchi2023}.

The same experiment also found a new resonance in the $^6$He + $p$ channel approximately 0.4 MeV above the $3/2^-3/2_1$ resonance, with no assignment of angular momentum and parity. The NCSMC calculation of Ref.~\cite{Vorabbi2019} predicted the $3/2^-1/2_3$ state in that region and it was thus considered as a possible match. In that energy region the coupled-partitions calculation predicts the $1/2^-1/2_3$ resonance, which contributes to both the cross sections of  $^6$Li($n,p)^6$He and $^6$He($p,n)^6$Li, and thus can be a candidate for the theoretical counterpart of the new measured resonance in Ref.~\cite{Dronchi2023}. 

Furthermore, the coupling of the partitions allowed us to calculate the charge-exchange reaction $^6$Li($n,p)^6$He cross section. The calculation approximately reproduces the reaction threshold and the overall shape of the cross section as a function of energy, but it does not reproduce the positions of peaks since the resonances are found at slightly different positions than experiment.

To improve the description of the shape of the cross section of the reaction $^6$Li($n,p)^6$He as well as of the elastic scattering of neutrons on $^6$Li, we adjusted the NCSM energies in the Hamiltonian kernel appearing in the NCSMC equations. After these adjustments, we were able to reproduce the positions of the two peaks in the cross section of the reaction $^6$Li($n,p)^6$He as well as the position of the $5/2^-1/2_2$ resonance in the cross section of the elastic scattering of neutrons. However, we were not able to simultaneously reproduce the experimental values of the elastic cross section at low energies.

We also calculated the cross sections of the reactions $^6$He($p,n)^6$Li and $^6$He($p,t)^4$He and differential cross sections of the elastic scattering of protons on $^6$He. These are to be compared with future results of an experiment planned at the Canada's particle accelerator center TRIUMF to probe resonances in $^7$Li above the $^6$He + $p$ threshold by impinging a beam of $^6$He onto a target with hydrogen thereby studying $^6$He + $p$ reactions in inverse kinematics.

We also calculated the differential cross section of the $^6$He($p,t)^4$He reaction. The calculated cross section reproduces the overall shape of the measured cross section, but is unable to reproduce the positions of the minima and maxima. Phenomenological adjustments help to better reproduce the positions of the minima and maxima, but the result overestimates the experimental values. It should be noted, however, that the energy $E_{\rm c.m.}=21.4$ MeV is beyond applicability of the current setup due to omitted channels which are also open at this high energy. No experimental data exist at lower energy for comparison. 

In the future we will carry out similar calculations for $^7$Be coupling the partitions $^4$He + $^3$He and $^6$Li + $p$. Our calculations can be improved by inclusion of three-nucleon forces, as already done, for example, in NCSMC calculation of the $^4$He($^3$He,$\gamma)^7$Be radiative capture~\cite{Atkinson2025}.

\begin{acknowledgments}
This work was supported by NSERC Grant No.~SAPIN-2022-00019 and by the U.S.~Department of Energy, Office of Science, Office of Nuclear Physics, under Work Proposal No.~SCW0498 and Award No.~DE-FG02-95ER40934. TRIUMF receives federal funding via a contribution agreement with the National Research Council of Canada. This work was prepared in part by LLNL under Contract No.~DE-AC52-07NA27344. Computing support came from an INCITE Award on the Frontier supercomputer of the Oak Ridge Leadership Computing Facility (OLCF) at ORNL, from the LLNL institutional Computing Grand Challenge program, and the Digital Research Alliance of Canada.
\end{acknowledgments}

\end{document}